\documentclass[
twocolumn,aps,pra,
 amsmath,amssymb]{revtex4-1}
\usepackage{mathrsfs}
\usepackage{graphicx} 
\usepackage{braket}
\usepackage{xcolor}
\begin{document}

\title{Unravelling the edge spectra of non-Hermitian Chern insulators}

\author{James Bartlett}
\affiliation{%
 Department of Physics and Astronomy, George Mason University, Fairfax, Virginia 22030, USA
}%
\author{Erhai Zhao}
\affiliation{%
 Department of Physics and Astronomy, George Mason University, Fairfax, Virginia 22030, USA
}%


\begin{abstract}
Non-Hermitian Chern insulators differ from their Hermitian cousins in one key aspect: their
edge spectra are incredibly rich and confounding. For example, even in the simple case where the
bulk spectrum consists of two bands with Chern number $\pm 1$, 
the edge spectrum in the slab geometry may have one or two edge states on both edges, or only at one of the edges, 
depending on the model parameters. 
This blatant violation of the familiar bulk-edge correspondence 
casts doubt on whether the bulk Chern number can still be a useful topological invariant, and demands 
a working theory that can predict and explain the myriad of edge spectra
from the bulk
Hamiltonian
to restore the bulk-edge correspondence.
We outline how such a theory can be set up to yield a thorough understanding of the
edge phase diagram based on the notion of the generalized Brillouin zone (GBZ) and the asymptotic
properties of block Toeplitz matrices. The procedure is illustrated by solving and comparing 
three non-Hermitian generalizations of the Qi-Wu-Zhang model, a canonical example of two-band 
Chern insulators. We find that, surprisingly, in many cases the phase boundaries and the number and location of 
the edge states can be obtained analytically.
Our analysis also reveals a non-Hermitian semimetal phase whose energy-momentum spectrum forms a continuous membrane
with the edge modes transversing the hole, or genus, of the membrane. Subtleties in defining the Chern number over GBZ, which in 
general is not a smooth manifold and may have singularities, are demonstrated using examples. 
The approach presented here can be generalized to more complicated models 
of non-Hermitian insulators or semimetals in two or three dimensions. 
\end{abstract}

\maketitle

\section{\label{sec:intro}Introduction}
In recent years, substantial progress has been made in characterizing the topological properties of non-Hermitian 
Hamiltonians describing non-interacting particles hopping on periodic lattices \cite{doi:10.1080/00018732.2021.1876991,RevModPhys.93.015005,https://doi.org/10.48550/arxiv.2205.10379,Ghatak_2019}.
Despite its apparent simplicity, many
aspects of the problem, especially in dimensions higher than one, still remain shrouded in mystery and lack 
the same level of completeness or clarity as the Hermitian topological phases of matter. 
To motivate our paper and to pinpoint the problem, we jump right to a concrete model. 
More detailed discussions of the background, including previous results that inspired and influenced our paper, will be given 
in Sec. VI.

\subsection{The Qi-Wu-Zhang model}

We are interested in the non-Hermitian generalizations of Chern insulators in two dimensions (2D).
A simple example of Hermitian Chern insulators is a two-band model introduced by Qi, \textit{et al.} on a square lattice \cite{PhysRevB.74.085308}.
Its Hamiltonian reads 
\begin{eqnarray}
H_{0}(\mathbf{k})&=&\mathbf{d}(\mathbf{k})\cdot \boldsymbol{\sigma} \label{QWZ}\\
&=&\sin k_{x}\sigma_{x}+\sin k_{y}\sigma_{y}
+(m-\cos k_{x}-\cos k_{y})\sigma_{z}, \nonumber
\end{eqnarray}
where $\mathbf{k}=(k_x,k_y)$ is the crystal momentum, $\boldsymbol{\sigma}=(\sigma_{x},\sigma_{y},\sigma_{z})$ denotes the Pauli matrices to describe the two orbital degrees of freedom (pseudo-spin 1/2), and the momentum-dependent ``magnetic field" $\mathbf{d}=(d_x,d_y,d_z)$ with $d_x=\sin k_x$, $d_y=\sin k_y$, $d_z=m-\cos k_{x}-\cos k_{y}$. The tuning parameter $m$ is real and plays the role of Dirac mass to dictate the energy gap. Note that the energy is always measured in units of the nearest-neighbor hopping $t$ which we have set to be 1. 
The Qi-Wu-Zhang model Eq. \eqref{QWZ}
has the virtue of being mathematically elegant with a clean-cut phase diagram  \cite{PhysRevB.74.085308}: For $|m|<2$, the system is topologically nontrivial with the Chern numbers of the two bands being $\pm 1$, the energy gap closes when $m=\pm 2$, and the system becomes topologically trivial for $|m|>2$. From the Chern numbers of the bands, we immediately know that there is one chiral edge mode inside the energy gap
for $|m|<2$.

\subsection{Three non-Hermitian generalizations}
\begin{figure*}[htpb!]
    \centering
 \includegraphics[scale=0.95]{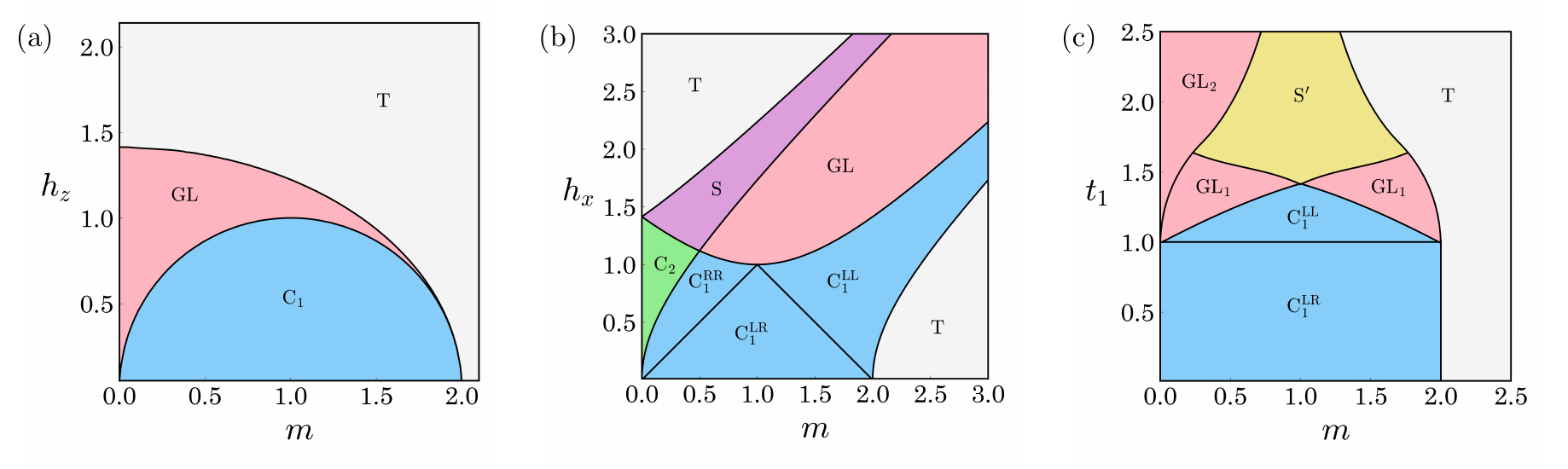}
    \caption{\label{fig:phase_diagrams} Summary of main results: A myriad of phases of non-Hermitian Chern insulators in slab geometry, $x\in [0, L]$ in the limit of large $L$ with periodic boundary conditions along $y$.
    The three panels (a)-(c) show the phase diagram for the generalized Qi-Wu-Zhang model $H_1$, $H_2$, and $H_3$, respectively. For model $H_1$ in (a), there are three phases indicated by color: the gapped topological phase with Chern number $\pm 1$ (C$_1$), the gapless phase (GL), and the trivial band insulator (T). This model is discussed in Sec. III. For $H_2$ in (b), the C$_1$ phase is further partitioned into three regions, where the subscripts such as LR describe the localization of the edge modes. There is also a gapped phase C$_2$ with Chern number $\pm 2$, and a topological gapless phase S. Both are unexpected from the bulk phase diagram.  See Sec. IV for details. 
    For model $H_3$ in (c), the C$_1$ phase has two regions with distinct edge behaviors, and a topological gapless phase S$'$ (Sec. V).}
\end{figure*}

A few different non-Hermitian generalizations of the Qi-Wu-Zhang model have been considered in the literature. Below, we will analyze and compare 
three examples.  
They are obtained by adding one extra term
to the Qi-Wu-Zhang model $H_0$, resulting {in} increasing complexity in the edge spectra. 
The first model is
\begin{equation}
H_1=H_0+i h_z\sigma_{z}. \label{H1}
\end{equation}
The constant $h_z$ term introduces an imaginary part to the ``magnetic field" $\mathbf{d}$ by replacing $d_z\rightarrow d_z+ih_z$ while retaining $d_{x,y}$ in $H_0$. In Ref. \cite{PhysRevLett.118.045701}, a three-dimensional generalization (with $k_z$) of this model was used to discuss Weyl exceptional rings. This model will be analyzed in Sec. III, and its phase diagram 
in the slab geometry (e.g., with open boundaries in the $x$ direction, $x\in [0,L]$, and periodic boundary condition along $y$)
is summarized in Fig. \ref{fig:phase_diagrams}(a).  We use $H_1$ as a warm-up example to set the stage for 
models $H_{2}$ and $H_3$ below, which feature much more complicated phase diagrams.
 
The second model is similar to $H_1$, but with the non-Hermitian term applied to $\sigma_x$ instead:
\begin{equation}
H_2=H_0+i h_x\sigma_{x}.
\end{equation}
This model has been investigated by Kawabata, \textit{et al.} in Ref. \cite{PhysRevB.98.165148} to illustrate the breakdown of
bulk edge correspondence in non-Hermitian Chern insulators. These authors obtained the phase diagram of $H_2$
in the slab geometry by numerical diagonalization.
Our objective here is to formulate 
an analytical theory to predict all the phase boundaries based on the notion of GBZ, without
resorting to numerical diagonalization of finite size systems with boundaries. This is done in Sec. IV,
and the analytical result for $L\rightarrow \infty$ is summarized in Fig. \ref{fig:phase_diagrams}(b). Note that
we label the various phases differently from Ref.  \cite{PhysRevB.98.165148}, for reasons to be elaborated on in Sec. IV.

The third model is defined as
\begin{equation}
H_{3}=H_0+it_{1}\sin k_{x}\sigma_{z} \label{H3}.
\end{equation}
A more general version of this model with an extra term involving $\sin k_y$ was considered in Ref. \cite{Zhang2022} as an example of 
the non-Hermitian skin effect. The phase diagram of $H_3$ in the slab geometry, however, remains unexplored to our best knowledge.
We solve this model in Sec. V and the resulting slab phase diagram is shown in Fig. \ref{fig:phase_diagrams}(c). 

\subsection{New phases in slab geometry}

The slab phase diagrams in Figs. \ref{fig:phase_diagrams}(b) and \ref{fig:phase_diagrams}(c) exhibit a few striking features when viewed alongside the corresponding bulk (i.e., with periodic boundary conditions in both $x$ and $y$ directions) phase diagrams. All three generalized Qi-Wu-Zhang models above bear the form
\begin{equation}
H_{\text{bulk}}(\mathbf{k})=\mathbf{D}(\mathbf{k})\cdot \boldsymbol{\sigma}, \label{hD}
\end{equation}
where the vector $\mathbf{D}$ depends on $\mathbf{k}$ and is in general complex. For example, for $H_3$ we have
$D_x=d_x$, $D_y=d_y$, $D_z=d_z+it_{1}\sin k_{x}$. Its bulk spectrum is simply
\begin{equation}
E_{\text{bulk}}(\mathbf{k})=\pm \sqrt{D_x^2+D_y^2+D_z^2},\label{ED}
\end{equation}
with $\mathbf{k}$ confined within the Brillouin zone (BZ). When the spectrum on the complex $E$ plane has
a well-defined line gap, one can compute the Chern number of each band from its biorthogonal eigenstates.
But the knowledge of $E_{\text{bulk}}(\mathbf{k}\in \mathrm{BZ})$ offers little help in comprehending the corresponding 
slab phase diagram for the case of $H_2$  or $H_3$.

Take $H_2$ for example. As highlighted in Fig. 1 of Ref. \cite{PhysRevB.98.165148}, 
its bulk phase diagram is partitioned by equally spaced diagonal lines on the $(m,h_x)$ plane. 
Three gapped phases with Chern number $C=0,\pm 1$, respectively, have the shape of a perfect diamond 
with side length $\sqrt{2}$. In contrast, the slab phase diagram of $H_2$ in Fig. \ref{fig:phase_diagrams}(b) is rather different.
Most striking is the appearance of a gapped phase C$_2$ (labelled as $N_{L/R}=2$ in Ref. \cite{PhysRevB.98.165148}) that 
has two edge modes on both the left ($x=0$) and right edge ($x=L$). This phase is unexpected, seemingly popped out from nowhere, 
because there is no bulk phase with Chern number $C=\pm 2$. 
The second notable feature is the emergence of two phases C$_{1}^{\mathrm{LL}}$ and 
C$_{1}^{\mathrm{RR}}$ (labelled as $N_{L}=2$ and $N_R=2$ in Ref. \cite{PhysRevB.98.165148}) 
with two edge modes localized at only one of the edges, which is impossible for Hermitian Chern insulators. 
It is obvious that these features cannot be inferred from $E_{\text{bulk}}(\mathbf{k}\in \mathrm{BZ})$,
and the familiar bulk-edge correspondence breaks down.

These observations beg the following questions. Can one predict
 when these caprice edge modes decide to switch sides, e.g., relocate from the left edge to the right edge as parameters
 $m$ and $h_x$ are varied? Moreover, what determines the curved phase boundaries of these phases?
The goal of our paper is to address these questions to achieve a more refined understanding of $H_2$. For instance,
we will prove in Sec. IV that all the phase boundaries of $H_2$ shown in Fig. \ref{fig:phase_diagrams}(b) are actually given by a set of
simple analytical curves. In Sec. V, we apply the theoretical technique developed 
to analyze the even more challenging case of $H_3$.

\subsection{Strategy to characterize the new phases}

Our strategy to comprehend non-Hermitian Chern insulators in the slab geometry
is built upon a few techniques developed earlier for one-dimensional (1D) non-Hermitian Hamiltonians.
A key idea is to 
include localized (non-Bloch) states besides the familiar Bloch waves by allowing the wave vector to take complex values. 
This is motivated in part by the non-Hermitian skin effect, i.e., the emergence of extensive number of eigenstates
localized at the open boundaries, e.g., at $x=0$ and/or $L$. For a given 1D tight-binding Hamiltonian $H(k_x)$, by replacing
$e^{ik_x}$ with a complex number $\beta$, one obtains an analytically continued Hamiltonian 
$\mathcal{H}(\beta)$:
\begin{equation}
H(k_x)\rightarrow \mathcal{H}(\beta).
\end{equation}
Its eigenvalues $E(\beta)$ will reproduce the open-boundary spectrum in the thermodynamic limit of $L\rightarrow \infty$ if $\beta$ is restricted to a closed curve on the complex plane known as the generalized Brillouin zone (GBZ):
\begin{equation}
\beta \in \mathrm{GBZ}. \label{gbz}
\end{equation}
In the context of 1D non-Hermitian band insulators, the concepts of ``non-Bloch" band theory and GBZ were first proposed in Ref. \cite{PhysRevLett.121.086803}; the correct definition of GBZ for the general case was given in Ref. \cite{PhysRevLett.123.066404}.  Once the GBZ is determined, one can define topological invariants such as the winding number. It was shown that the phase boundaries obtained from $\mathcal{H}(\beta\in \mathrm{GBZ})$ match those from numerical diagonalization of finite-size systems with open boundaries. In this way, the bulk-boundary correspondence is restored
by introducing GBZ for 1D non-Hermitian Hamiltonians.
 
At first, one might expect that this approach can be generalized trivially to two dimensions to describe non-Hermitian Chern insulators. Consider, for instance, $H_2$ or $H_3$ in the slab geometry with open boundaries at $x=0,L$ and periodic along $y$. One can follow the 1D recipe  by the replacement 
\begin{equation}
e^{ik_x} \rightarrow \beta 
\end{equation}
to construct an analytically continued Hamiltonian,
\begin{equation}
H(k_x,k_y)\rightarrow \mathcal{H}(\beta,k_y) ,
\end{equation}
where $k_y$ is a good quantum number. ${\mathcal{H}}(\beta,k_y)$ can be viewed as a 1D Hamiltonian with parameter $k_y$.
For each given $k_y$, one may compute the corresponding GBZ curve for $\mathcal{H}(\beta,k_y)$:
\begin{equation}
\beta \in \mathrm{GBZ}({k_y}).
\end{equation}
In principle, these $k_y$-dependent GBZ curves will congregate into a 2D surface in the space of $(\mathrm{Re}\beta, \mathrm{Im}\beta, k_y)$. Let us call this 2D surface the GBZ surface, or GBZ$_s$, to differentiate it from the 1D GBZ curve:
\begin{equation}
\mathrm{GBZ}_s =\cup_{k_y} \mathrm{GBZ}({k_y}). \label{GBZS}
\end{equation}
It reduces to the two-torus BZ if the Hamiltonian is Hermitian. One can proceed to define Chern numbers on GBZ$_s$ and use them to characterize each phase of $\mathcal{H}(\beta,k_y)$. If everything works out as expected, the resulting phase diagram should agree with the numerical diagonalization of large-size systems.

 \begin{figure}[htpb!]
    \centering
    \includegraphics[scale=1]{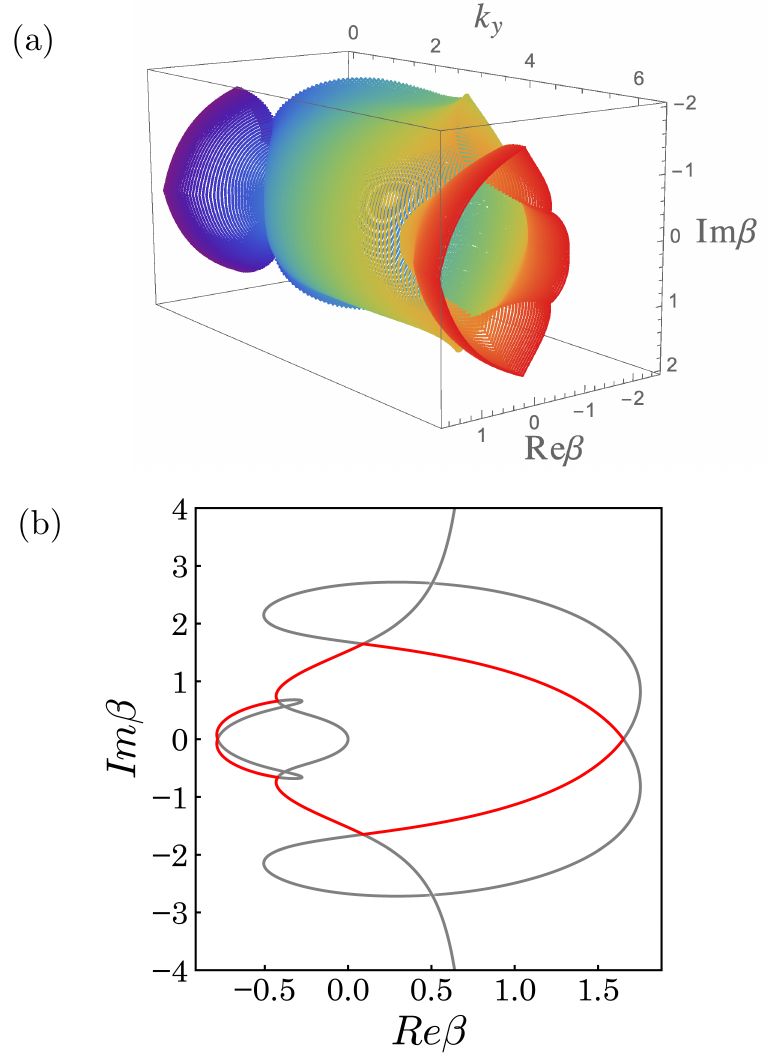}
    \caption{\label{fig:H3_GBZ} (a) An example of the generalized Brillouin zone surface (GBZ$_s$) for model 
    $H_3$. It is a three-dimensional surface in the space $(\mathrm{Re}\beta, \mathrm{Im}\beta, k_y)$,
   continuous but not necessarily smooth. Different values of $k_y\in [0,2\pi]$ are colored coded.
   (b) The cross section of the GBZ$_s$ at $k_y=0$. The red curve is the GBZ showing numerous cusps. 
    It consists of arc segments belonging to multiple auxiliary GBZ curves (in grey) which intersect with each other. 
    Thus, {the} GBZ$_s$, in general is not a smooth manifold but an algebraic variety.
    The results are obtained numerically by following the algorithm outlined in Sec. II.C.
    The parameters used are $m=0.6$, and $t_1=0.4$. 
    }
\end{figure}

In reality, carrying out this plan runs into difficulties. The GBZ surface is, in general, not a smooth compact manifold like the two-torus. This makes the definition and numerical computation of the Chern number challenging. Computing the GBZ curve  for 1D non-Hermitian Hamiltonians is a nontrivial task. A few powerful algorithms have been developed so far \cite{PhysRevLett.123.066404,PhysRevLett.125.226402,PhysRevB.105.045422}, and they rely on numerical solution of algebraic equations (e.g., finding the roots of polynomial equations or the intersections of two curves) to yield a collection of discrete data points for $\beta$ on the complex plane. With sufficient resolution, these data points coalesce into a curve which is believed to be continuous and closed, but not necessarily smooth. In fact, it often has sharp turns, or cusps. An example is the red curve in Fig. \ref{fig:H3_GBZ}(b). After running the algorithm for each $k_y\in [0,2\pi]$, the resulting GBZ$_s$ inherits these cusps, see the example shown in Fig. \ref{fig:H3_GBZ}(a).

To make matters worse, 
the resulting GBZ$_s$ sometimes features singularities as $k_y$ is varied. For example, each GBZ curve of $H_2$ is a circle of radius $r$ on the complex plane, but $r$ shrinks to zero or blows up to infinity at certain $k_y$ values as shown in Fig. 3. 
Thus, to evaluate the Chern number from the Berry connection for a general non-Hermitian Chern insulator, we have to settle for irregular mesh points on a rugged GBZ$_s$, and watch out { for} singularities, e.g. when the GBZ surface shrinks to a point. Note that, 
previously, in Ref. \cite{Gang-Feng_Guo}, the singularities of GBZ in 1D non-Hermitian models have been noted. Here, we focus on 2D models.
To circumvent these subtleties and crosscheck the Chern number calculation, we shall also pursue an alternative scheme to characterize the topological invariant of ${\mathcal{H}}(\beta, k_y)$ using the eigenvectors on the Bloch sphere.

From Secs, III-V, the three models $H_{1,2,3}$ are discussed, in turn, to illustrate the technical complexities and challenges in executing the strategy outlined above centering around $\mathcal{H}(\beta,k_y)$ and GBZ$_s$. In particular, we show how analytical solutions can be obtained for $H_2$ to yield a thorough understanding of the problem. By working through these three examples, we hope the reader can appreciate  the rich, nontrivial behaviors of non-Hermitian Chern insulators with open boundaries as highlighted in Fig. \ref{fig:phase_diagrams}.

\section{Computing the GBZ}

The concept of a GBZ plays a crucial role in our analysis of the non-Hermitian Chern insulators. 
In this section, we outline the technical procedures to compute the GBZ$_s$ for our two-band models. Based on existing
algorithms, we introduce a few tricks so the numerical task
is simplified and analytical results become possible. This leads to a rather detailed knowledge of how {the} GBZ varies with the 
 parameters such as $m$, $h_x$ or $t_1$, and $k_y$, including the development of singularities. 

\subsection{The algorithm}
The first step of the algorithm is to
analytically continue $H_{\text{bulk}}(k_x,k_y)$ by the replacement $e^{ik_x}\rightarrow \beta$.
Take model $H_2(k_x,k_y)$ as an example. After the replacement, $H_2$ becomes 
\begin{eqnarray}
\mathcal{H}_2(\beta,k_y)&=&\big(ih_x+\frac{\beta-\beta^{-1}}{2i}\big)\sigma_x +\sin k_y\sigma_y \nonumber \\
&&+  \big(m-\cos k_y-\frac{\beta+\beta^{-1}}{2}\big)\sigma_z. \label{H2beta}
 \end{eqnarray}
The two eigenvalues of $\mathcal{H}_2$ are $\pm E(\beta)$, where the $k_y$ dependence of $E$ has been suppressed for brevity. We will focus on their square,  which is a Laurent polynomial of complex variable $\beta$:
\begin{equation}
\epsilon(\beta)\equiv {E}^2(\beta)=\frac{a' \beta^2 + b'\beta +c'}{\beta}. \label{Epoly2}
\end{equation}
Here the coefficients 
\begin{eqnarray}
a'&=&-m+\cos k_y+h_x, \\
b'&=&m^2-2m\cos k_y+2-h_x^2,\\
c'&=&-m+\cos k_y- h_x.
\end{eqnarray}
For $\beta$ living on the unit circle with $|\beta|=1$, let $\beta=e^{ik_x}$ with $k_x\in [-\pi,\pi]$, then $\epsilon(\beta)$ becomes $\epsilon(k_x)$ to reproduce the bulk spectrum Eq. \eqref{ED}, i.e., the spectrum of $H_2$ with periodic boundary conditions.

To discuss the slab geometry with open boundaries and the corresponding edge modes, let us write $H_2$ in second quantized form: 
\begin{equation}
H_2=\sum_{n=1}^L\sum_{k_y} \left[ \psi^\dagger_nA\psi_n +  \psi^\dagger_nB\psi_{n+1} +  \psi^\dagger_{n+1}C\psi_{n}\right].\label{T}
\end{equation}
Here the good quantum number $k_y$ is the crystal momentum along $y$, 
$n$ is the unit cell index along $x$,
the creation operator $\psi^\dagger_n$ is a shorthand notation for the spinor $[\psi^\dagger_{n,\uparrow}(k_y),\psi^\dagger_{n,\downarrow}(k_y)]$, and $A$, $B$, $C$ are $2\times 2$ matrices:
\begin{eqnarray}
 A &=& [m-\cos k_y] \sigma_z + ih_x\sigma_x +\sin k_y\sigma_y, \label{Adef}\\
B &= &(-\sigma_z + i\sigma_x)/2,\\
C &=& B^\dagger.\label{Cdef}
\end{eqnarray}
Again, the $k_y$ dependence of $\psi_n$ and $A$ is suppressed for brevity.
In other words, $H_2$ is a block Toeplitz matrix:
\begin{equation}
\mathcal{T}=\left[
\begin{array}{ccccc}
A&  B & 0 &0  &.. \\
C&  A & B &0   & ..\\
0&  C & A  & B &.. \\
..&  .. & ..  & .. &.. 
\end{array}
\right]_{L\times L}. \label{Tmatrix}
\end{equation}
This form is particularly convenient for finding the dispersion and location of the edge states in Sec. IV.

We stress that the open spectrum (i.e., the set of eigenvalues of $\mathcal{T}$ for a system of size $L$ in the $x$ direction with open boundaries
at $x=0,L$) may bear little similarity with the bulk spectrum $E_{\text{bulk}}(k_x,k_y)$ in Eq. \eqref{ED}. Furthermore,  the open spectrum depends on the size $L$. And since $\mathcal{T}$ is non-Hermitian, its numerical diagonalization is prone to instabilities and large errors when $L$ is large. Some of these counterintuitive phenomena have been long noticed for 
non-Hermitian Toeplitz matrices (here we are dealing with block Toeplitz matrices). A simple example is when the three matrices $A$, $B$, and $C$ reduce to numbers with $A=0$. In this case, the bulk spectrum is an ellipse (a curve), while the open spectrum is a line segment within the ellipse.
The sensitive dependence of the spectra on the boundary conditions has been well recognized for non-Hermitian Hamiltonians. 

The spectra of $\mathcal{T}$ in the thermodynamic limit $L\rightarrow \infty$, save for a subset corresponding to the edge states, are 
called {continuum bands}. The eigenvalues of $\mathcal{T}$ congregate into continuum sets (e.g., lines) on the complex $E$ plane.
And the corresponding eigenstates may include localized states that do not belong to the bulk spectrum with periodic boundaries.
A key step in understanding the slab phase diagrams of non-Hermitian Chern insulators is to find the continuum bands.
Remarkably, this task can be reduced to an algebraic problem 
involving the analytically continued Hamiltonian $\mathcal{H}(\beta)$. The eigenvalue
problem of $\mathcal{H}(\beta)$ has the following generic form:
\begin{equation}
\frac{P_{p+q}(\beta,E)}{\beta^p} = 0. \label{polygen}
\end{equation}
Here $P_{p+q}$ denotes a polynomial of $\beta$ of degree $p+q$ (it is also a polynomial of $E$ of some other degree). For example, 
in Eq. \eqref{Epoly2} for $\mathcal{H}_2$, we have $p=1$ and $q=1$. For generalized Qi-Wu-Zhang models,
Eq. \eqref{polygen} can be further simplified to a form similar to Eq. \eqref{Epoly2},
\begin{equation}
\epsilon(\beta)= \frac{Q_{p+q}(\beta)}{\beta^p}, \label{Esqpoly}
\end{equation}
where $Q_{p+q}$ is a polynomial of $\beta$ of degree $p+q$ with coefficients independent
of $E$. Equation \eqref{Esqpoly} defines a mapping 
$\beta\rightarrow \epsilon$, i.e.,
from $\beta$ on the complex plane to $\epsilon=E^2$, which is also complex. Note that this is
a multiple-to-one mapping: For a given image $\epsilon$, we label its preimages by $\beta_i$
and order them by their magnitudes:
\begin{equation}
|\beta_1| \leq |\beta_2| \leq ...\leq |\beta_{p+q}|.
\end{equation}
For $E$ to lie within the continuum band, its preimages $\beta_p$ and $\beta_{p+1}$ must satisfy
the degeneracy condition:
\begin{equation}
|\beta_p| = |\beta_{p+1}|. \label{degencond}
\end{equation}
This key result was established in Refs. \cite{PhysRevLett.121.086803} and \cite{PhysRevLett.123.066404} for 1D non-Hermitian Hamiltonians. And in the context of 
Toeplitz matrices, it was first proved by Schmidt and Spitzer (see Theorem 1 of Ref. \cite{10.2307/24489115} and Refs. \cite{BeamRichardM.1991Taso,doi:10.1137/1.9780898717853} for a review). Solving Eq. \eqref{degencond} together with Eq. \eqref{Esqpoly} accomplishes two goals at once:
the set of $E$'s form the continuum band, and the union of the set $\beta_p$ and set $\beta_{p+1}$ gives the 
GBZ. 

\subsection{Circular GBZ for model $H_2$}

Let us apply the algorithm to $H_2$ to show its GBZ is a circle. 
Recall from Eq. \eqref{Epoly2}, $\epsilon(\beta)=a'\beta + b'+ c'/\beta$, so $p=1$ and each $\epsilon$ only has
two preimages, $\beta_1$ and $\beta_2$. The degeneracy condition requires them to have equal magnitude,
therefore we can follow the parametrization scheme of Ref. \cite{PhysRevLett.123.066404} to write
\begin{equation}
\beta_2 = \beta_1 e^{i \theta}.
\end{equation}
Next, using $\epsilon(\beta_1)=\epsilon(\beta_2=\beta_1e^{i\theta})$, we find
\begin{equation}
\beta^2_1 =\frac{c'}{a'} e^{-i \theta}.
\end{equation}
Thus the GBZ for $\mathcal{H}_2$ is a circle with radius
\begin{equation}
r_2 (k_y)=\left| \frac{m-\cos k_y+h_x}{m-\cos k_y-h_x}\right|^{1/2}. \label{radiusr2}
\end{equation}
We stress that  working with $\epsilon=E^2$ is 
a simple yet crucial trick to render the problem analytically tractable.

\begin{figure}[htpb!]
    \centering
    \includegraphics[width=0.4\textwidth]{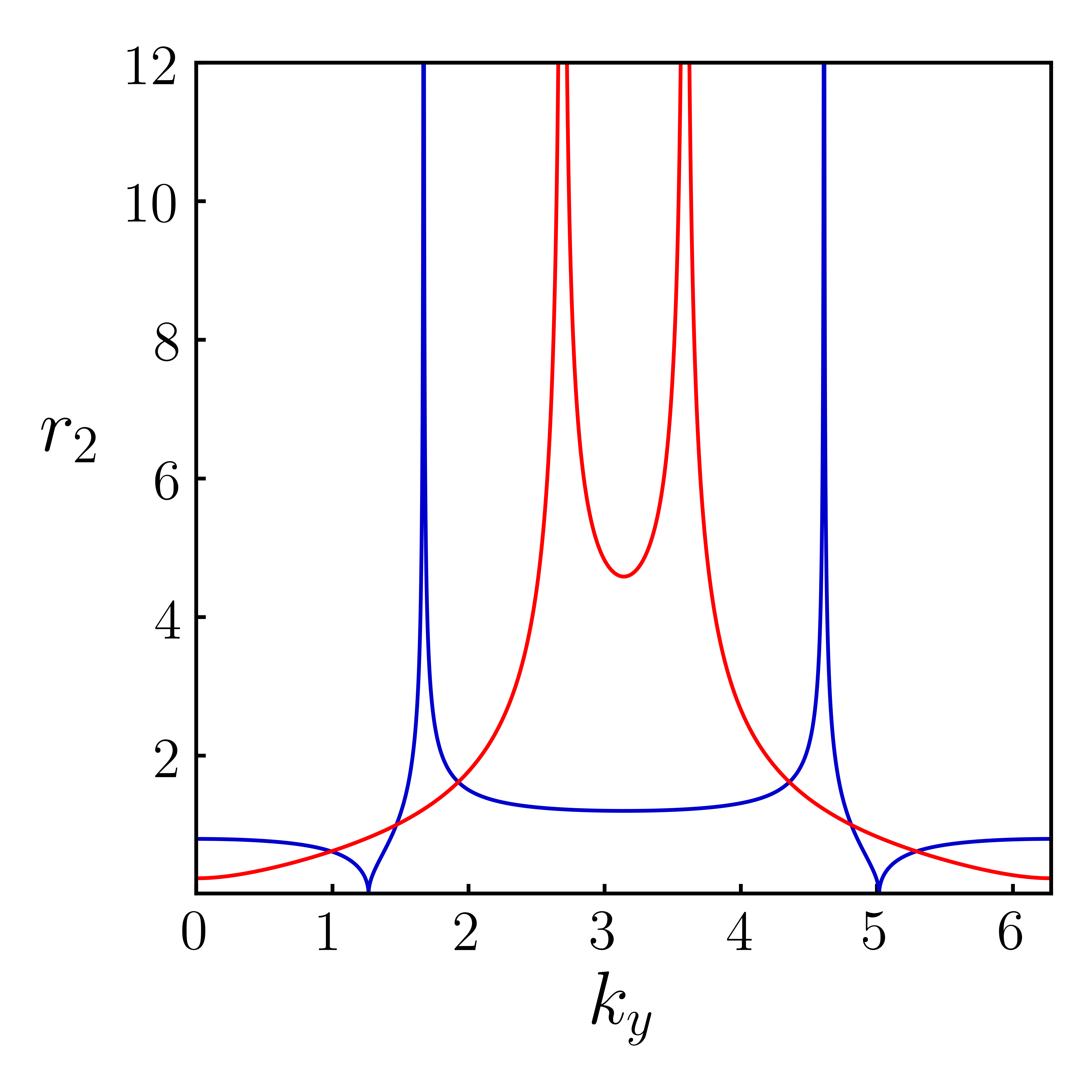}
    \caption{\label{fig:H2_GBZ_r} {Singularities in the circular GBZ of model $H_2$. The radius $r_2$, given by Eq. \eqref{radiusr2}, 
    is plotted as a function of $k_y\in [0, 2\pi]$ for $m=0.1$, $h_x=0.2$ (in blue), and for $m=0.1$, $h_x=1$ (in red). 
    It shrinks to zero when $m-\cos k_y+h_x=0$, and blows up to infinity when
    $m-\cos k_y-h_x=0$.}}
\end{figure}

Figure \ref{fig:H2_GBZ_r} shows two examples of {the} GBZ radius varying with $k_y$. It clearly illustrates the pinching of the GBZ$_s$ where $r_2$ vanishes,
as well as the divergence of $r_2$ at certain $k_y$ values dependent on the parameter $m$ and $h_x$.

\subsection{GBZ for model $H_3$}

Next we apply the algorithm to the analytically continued model $H_3$, which has the form
\begin{eqnarray}
\mathcal{H}_3(\beta,k_y)&=&\frac{\beta-\beta^{-1}}{2i}\sigma_x +\sin k_y\sigma_y  \label{H3beta} \\
&+&  \big(m-\cos k_y+ \frac{1-t_1}{2}\beta -\frac{t_1+1}{2}\beta^{-1}\big)\sigma_z.\nonumber 
 \end{eqnarray}
Its eigenvalue square is the Laurent polynomial
\begin{equation}
\epsilon (\beta) \equiv {E}^2(\beta)=a\beta^2 + b/\beta^{2}+c\beta+d/\beta+f , \label{atof}
\end{equation} 
with 
\begin{eqnarray}
a&=&t_1(t_1-2)/4, \\
b&=&t_1(t_1+2)/4,\\
c&=&+ (m-\cos k_y)(t_1-1),\\
d&=&-(m-\cos k_y)(t_1+1),\\
f&=&(m-\cos k_y)^2+\sin^2 k_y-t_1^2/2+1. \label{f_def_3}
\end{eqnarray}
Comparing with the general form Eq. \eqref{Esqpoly}, we find that, in this case, $p=q=2$, so the degeneracy condition becomes 
\begin{equation}
|\beta_2| = |\beta_{3}|.  \label{degen23}
\end{equation} 
This motivates the parametrization $\beta_3=\beta_2e^{i\theta}$. To find the GBZ,
we need to solve the quartic equation
\begin{equation}
\epsilon(\beta_2)=\epsilon(\beta_2 e^{i\theta}) \label{theta-param}
\end{equation} 
for $\beta_2$ for all possible values of $\theta$. 

Before attempting a general solution, let us first consider a special case $m=1$ and $k_y=0$, so $c=d=0$. Then the Laurent polynomial Eq. \eqref{atof} simplifies to
$\epsilon(\beta)=a\beta^2 + b/\beta^{2}+f$, and Eq. \eqref{theta-param} 
can be solved by hand. After a little algebra, we find
 \begin{equation}
\beta^4_2 =\frac{t_1+2}{t_1-2} e^{-i 2\theta}.
\end{equation}
Thus, the GBZ for $m=1$ and $k_y=0$ is a circle of radius
 \begin{equation}
r_3 =\left| \frac{t_1+2}{t_1-2} \right|^{1/4}. \label{radius3}
\end{equation}

For general values of $(m,t_1,k_y)$,
we choose to solve Eq. \eqref{theta-param} numerically to find its four solutions $\eta_i$, $i=1,2,3,4$. For each solution $\eta_i$, 
we compute its image $\epsilon^*=\epsilon(\eta_i)$, and find all four preimages of $\epsilon^*$ and sort them into 
\begin{equation}
|\xi_1| \leq |\xi_2| \leq \xi_3| \leq |\xi_4|. \label{orderxi}
\end{equation}
If  $|\xi_2|= |\xi_3|$ and $\eta_i=\xi_2$ or $\xi_3$,  
we conclude that $\eta_i$ and $\eta_ie^{i\theta}$ satisfy the condition Eq. \eqref{degen23}
and therefore belong to the GBZ.
Repeating this procedure for a discrete grid of $\theta$ values within the interval $[0,2\pi]$ will produce a set of data points to form the GBZ curve
[e.g., the red curve in Fig. \ref{fig:H3_GBZ}(b)].

It is not immediately obvious that the GBZ obtained this way is guaranteed to be a connected, closed curve. 
To gain a better understanding, it is useful to examine all the solutions of Eq. \eqref{theta-param}, including those that do not 
meet the criterion Eq. \eqref{degen23} and thus do not belong to the GBZ. As the solutions of a polynomial equation, they
forms continuous curves on the complex $\beta$ plane which are called the auxiliary GBZ by Ref. \cite{PhysRevLett.125.226402}. For example, some of the 
solutions satisfy $|\beta_j|=|\beta_{j+1}|$, with $j\neq q$.
These curves may intersect, and GBZ is nothing but a subset of the auxiliary GBZ, consisting of 
arcs {\it connected} to each other at these intersection points.
Figure \ref{fig:H3_GBZ} shows the computed GBZ (in red) and other auxiliary GBZ for parameters $m=0.6$, $t_1=0.4$ with $k_y=0$. 

It is clear from the discussion above that {the} GBZ, unlike the familiar BZ, 
 is not necessarily a smooth manifold and may feature singular points. More precisely, it should be called an {\it algebraic variety}, as
it is derived from solutions to polynomial equations.

 \subsection{GBZ from self-intersection}

The recipe outlined in the preceding subsections works very well in
 tracing out smooth GBZ curves, e.g., approximately of elliptical shape. Its performance suffers, however, when the
GBZ contains segments going along the radial direction, which can be easily missed if the mesh grid of 
$\theta$ is not fine enough. Thus, it is useful to develop an alternative method that can find points on {the} GBZ
at a given radius $\rho$ on the complex $\beta$ plane. An ingenious algorithm of this type was proposed in Ref. \cite{PhysRevB.105.045422}
based on the self-intersection and winding of the image $\epsilon(\beta)$. 
Below, we show how it can be adapted to $\mathcal{H}_3$. Readers who are not interested in
these technical details can skip to Sec. III.

Let $\mathscr{C}_\rho$ be a circle of given radius $\rho$ on the complex plane. As $\beta$ varies along $\mathscr{C}_\rho$
to complete a cycle, its image $\epsilon(\beta)$ traces out a closed curve $\Gamma_\rho$ on the complex plane of $\epsilon$:
\begin{equation}
\Gamma_\rho =\left\{ \epsilon (\beta\in \mathscr{C}_\rho) \right\}.
\end{equation}
Thus, for two distinct points $\beta_i$ and $\beta_j$ on $\mathscr{C}_\rho$ to map to the same image $\epsilon_s\in \Gamma_\rho$,
\begin{equation}
\epsilon(\beta_i\in \mathscr{C}_\rho) = \epsilon(\beta_j\in \mathscr{C}_\rho) = \epsilon_s, \label{degencross}
\end{equation}
$\epsilon_s$ must be a self-intersection points of the curve $\Gamma_\rho$. For our problem, 
we observe that the location of these points are mirror symmetric with respect to the real axis,
 because all coefficients $a$ to $f$  in Eq. \eqref{atof} are real. 

Plotting the curve $\Gamma_\rho$ reveals that it is, in general, very complicated.
One may take a purely numerical approach to find its intersection points. But it is time consuming 
(we must repeat the calculation for different $\rho$'s and different parameters such as $k_y$) and 
requires fine-tuning for different parameters. It turns out that with some effort all the self-intersection points for model $\mathcal{H}_3$
can be found analytically as follows.
For a given radius $\rho$, let us parametrize $\beta=\rho e^{i\theta}$ and separate $\epsilon$ into real and imaginary parts, 
$\epsilon(\beta=\rho e^{i\theta})=x(\theta)+iy(\theta)$. Then Eq. \eqref{atof} becomes two equations, 
\begin{align}
x(\theta)=a_+\cos 2\theta + c_+\cos \theta, \\
y(\theta)=a_-\sin 2\theta + c_-\sin \theta,
\end{align}
with the shorthand notation
\begin{align}
a_\pm&=a\rho^2\pm {b}{\rho^{-2}}, \\
c_\pm&=c\rho\pm {d}{\rho^{-1}}.
\end{align}
According to Eq. \eqref{degencross}, a self-intersection point of $\Gamma_\rho$ corresponds to a solution to the equation set
\begin{align}
x(\theta)=x(\theta')\label{xx}, \\
y(\theta)=y(\theta') \label{yy},
\end{align}
with $\theta\neq \theta'$. These trigonometric equations can be converted into polynomial form by introducing
\begin{equation}
u=\cos\theta,\;\;\; v= \cos\theta' ,
\end{equation}
and applying trig identities. For example, Eq. \eqref{xx} for $x$ reduces to
\begin{equation}
 u+v=-c_+/2a_+ \label{u+v}
\end{equation}
after factoring out $(u-v)$. Eq. \eqref{yy} for $y$ is more involved. One can square it to obtain a quartic equation for $u$ and $v$ using $\sin^2\theta=1-u^2$. Luckily, we can factor out $(u-v)$ again, and evoke Eq. \eqref{u+v} to reduce it to a quadratic equation for $u$,
\begin{equation}
{a}_u u^2 + {b}_u u + {c}_u=0, \label{u2}
\end{equation}
where the coefficients have lengthy expressions
\begin{align*}
& \eta = -c_+/2a_+, \\
&{a}_u=8a_-^2\eta+4a_-c_- , \\
&{b}_u=-8a_-^2\eta^2-4a_-c_-\eta ,\\
&{c}_u=4a_-^2\eta^3+4a_-c_-\eta^2+(c_-^2-4a_-^2)\eta-4a_-c_- .
\end{align*}
The quadratic Eq. \eqref{u2} yields a pair of solutions $u_{\pm}$.
Another independent solution of Eq. \eqref{yy} corresponds to $y(\theta)=y(\theta')=0$, 
leading to
\begin{equation}
u_3=-c_-/2a_- ,
\end{equation}
with the corresponding intersection point lying on the real axis. 
For each solution of $u$, we can work backwards to find $\theta=\arccos (u)$, $v$, $\theta'$,
and $\epsilon_s=x(\theta)+iy(\theta)$. 

In some special cases, the self-intersection points coincide and merge into a single point. This corresponds to having three $\beta$'s on the circle $\mathscr{C}_\rho$ that map to the same value of $\epsilon$. In Ref. \cite{PhysRevB.105.045422} this is called three-bifurcation point. Let the three $\beta$'s be $\beta_1=\rho_c$, $\beta_2=\rho_ce^{i\theta}$, and $\beta_3=\rho_ce^{-i\theta}$. They are the solutions of the quartic equation:
\begin{equation}
a\beta^4+c\beta^3-\epsilon \beta^2+d\beta +b=0.
\end{equation}
Using Vieta's formulas, after eliminating $\theta$, we find $\rho_c$ is the solution of a high order equation
\begin{equation}
a^2 \rho_c^8 +ad \rho_c^5 -cb \rho_c^3 -b^2=0, \label{order7}
\end{equation}
which can be solved numerically, e.g., using MATHEMATICA. Once $\rho_c$ is known, $\theta$ can be found via
\begin{equation}
\cos\theta = -\frac{1}{2}(\rho_c^4a/b+\rho_cd/b+1).
\end{equation}
This example illustrates the modest algebraic price one has to pay to understand the continuum bands
of non-Hermitian Chern insulators. To summarize, the self-intersection points $\epsilon_s$ can be 
found analytically from the values of $a,b,c,d$, and $\rho$, except for solving Eq. \eqref{order7} for the special case of higher-order bifurcation points. 

Not all the self-intersection points $\epsilon_s$ found above belong to the continuum band. Reference \cite{PhysRevB.105.045422} established a 
qualifying criterion: The neighborhood of  $\epsilon_s$ is divided into four regions by the two intersecting lines at $\epsilon_s$;
with respect to a chosen point $\epsilon_w$ in one of these regions and away from $\Gamma_\rho$, the winding number of 
the curve $\Gamma_\rho$ defined by
\begin{equation}
W (\epsilon_w)= \frac{1}{2\pi}\int_{\Gamma_\rho} dz \mathrm{Arg}(z-\epsilon_w)
\end{equation}
should be $+1,0,-1,0$ respectively. (The patterns of the winding number near a higher-order birfurcation points 
are more complicated and discussed in Ref. \cite{PhysRevB.105.045422}). The winding number is easy to evaluate numerically 
with the help of the argument principle,
\begin{equation}
W(\epsilon_w) = N_w - p ,
\end{equation}
where $p=2$ and $N_w$ is the number of the preimages of $\epsilon_w$ residing inside the circle $\mathscr{C}_\rho$. 
For those qualified self-intersection points with the right set of winding numbers, we collect their preimages $\rho e^{i\theta}$
on the circle $\mathscr{C}_\rho$ as a subset of GBZ. By changing the radius $\rho$ and repeating the procedure, one obtains
the whole GBZ curve.

What about those rejected $\epsilon_s$ with the ``wrong" winding number patterns? Their preimages are 
nothing but the auxiliary GBZ. The self-intersection method described here is complementary to the scheme
given in the preceding subsections.  It excels at resolving the cusps where the other method struggles. 
We have checked that these two methods agree with each other and 
produce the same GBZ as well as the auxiliary GBZ.
 
\section{Exceptional ring of model 1}
Some non-Hermitian Chern insulators are adiabatically connected to the familiar Hermitian Chern insulators.  
One example is the model $H_1$ defined in Eq. \eqref{H1} by replacing
$m$ with a complex Dirac mass $m+ih_z$ in the Qi-Wu-Zhang model $H_0$.
In this case, the bulk-edge correspondence holds 
as usual, and there is no need for introducing the notion of {the} GBZ. The other two models $H_{2,3}$, in comparison, 
will not be so cooperative. 
Model $H_1$ provides a nice geometric picture of the band topology in terms of the $\mathbf{d}$ vectors.
Here we show that the phase diagram of $H_1$ on the $(m,h_z)$ plane [Fig. \ref{fig:phase_diagrams}(a)] 
can be understood quantitatively by analyzing the
 the singularity of $H_1$ in the space of $\mathbf{d}$. 
This viewpoint based on the $\mathbf{d}$ vectors was advocated in Ref. \cite{PhysRevB.100.075403}
for a more complicated model.

As $k_x$ and $k_y$ vary throughout the BZ, the vector $\mathbf{d}(k_x,k_y)$ defined 
in Eq. (1) traces out a closed surface 
in the space of $(d_x,d_y,d_z)$. The surface is mirror symmetric with respect to the plane $d_z=m$, 
where it becomes pinched along the diagonal lines $|d_x|=|d_y|\in [0,1]$. 
It is useful to imagine the upper half of the surface as a bloated tent of height 2
with its bottom stitched together along two lines on the ground.
The eigenvalues of $H_1(k_x,k_y)$ will vanish when 
\begin{equation}
d_x^2+d_y^2+(d_z+ih_z)^2= 0 .
\end{equation}
And the singularity here is an exceptional point. 
Separating the real and imaginary parts, we find the condition becomes
\begin{equation}
d_z=0,\;\;\;\;d_x^2+d_y^2=h_z^2. \label{ring}
\end{equation}
This defines a ring of radius $h_z$ on the plane of $d_z=0$. 
We will call it the exceptional ring (Ref. \cite{PhysRevB.100.075403}
uses the more generic name ``singularity ring"). 
In the limit of $h_z=0$, the Qi-Wu-Zhang model is recovered,
and the ring shrinks to a point at the origin which, since the work of Berry \cite{doi:10.1098/rspa.1984.0023}, 
is often called a magnetic monopole
carrying magnetic charge. In this sense, the ring here is {\it a ring of magnetic charge}.

 \begin{figure}[htpb!]
    \centering
    \includegraphics[width=0.46\textwidth]{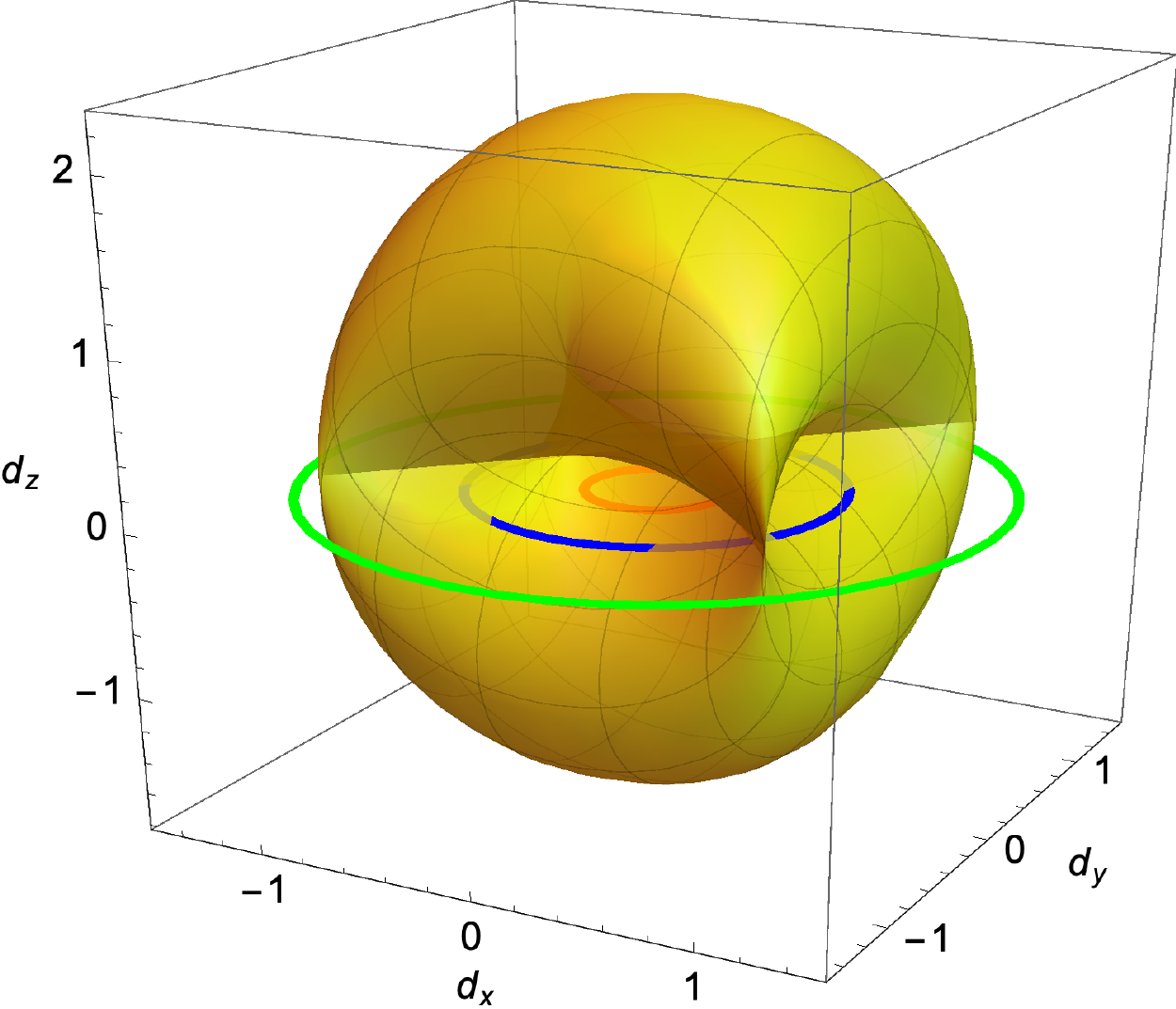}
    \caption{\label{fig:H1_tent} The ring and the tent: Geometric visualization of the $\mathbf{d}$ vector (golden surface, the tent)
    and the exceptional ring, Eqs. \eqref{ring}, for model $H_1$. At fixed $m=0.25$, for $h_x=0.3$, the exceptional ring 
    (red) resides inside the tent, the system is within the topological phase C$_1$. For $h_x=0.8$, the ring
    (blue) intersects the tent surface, the system is gapless. For $h_x=1.5$, the ring is outside the tent,
    the system is gapped but topologically trivial. The base of the tent is at $m=0.25$. 
    }
\end{figure}

Now the phase diagram of model $H_1$ can be worked out from the geometries of the tent
(centered at $d_z=m$ with overall height $2|m|$) and the exceptional ring 
(centered around $d_z=0$ with radius $h_z$), see Fig. \ref{fig:H1_tent}. 
When the exceptional ring lives inside/outside
the tent, the system is {a} topologically nontrivial/trivial insulator; when the
ring intersects the tent, the spectrum is gapless.
Figure \ref{fig:phase_diagrams}(a) shows the phase diagram of $H_1$ where the two gapped phases are separated
by the gapless region. By examining the cross section of the tent surface
with the $d_z=0$ plane and how it touches the ring, one can determine the phase boundaries. 
For example, for $m=1$
the lower critical point is at $h_z=1$ and the upper critical point is at $h_z=\sqrt{3/2}$.
At $m=0$, the transition to the trivial gapped phase occurs at $h_z=2$.
And the gap closes at $m=2$ and $h_z=0$. One can check that the edge states have real energy,
and the bulk-edge correspondence holds for $H_1$.

\section{Analytical theory of Model 2}
In this section, we revisit the phases and edge modes of $H_2$, which has been investigated numerically in Ref. \cite{PhysRevB.98.165148}. 
As discussed in the Introduction, our goal here is to achieve an analytical understanding. To this end,
we shall restrict our focus to the first quadrant of the $(m,h_x)$ plane with $m,h_x>0$. The phase diagram in other quadrants can 
be obtained by using symmetry. In particular, we establish the following eight theorems. 

{\bf Theorem 1}. The GBZ for model $\mathcal{H}_2(\beta,k_y)$ defined in Eq. \eqref{H2beta} is a circle of radius $r_2$ given by Eq. \eqref{radiusr2}. 

{\bf Theorem 2}. The band structure of $\mathcal{H}_2(\beta\in \mathrm{GBZ},k_y)$ defines three topologically nontrivial phases 
with robust edge states. Phase C$_1$ (C$_2$) has line gap and band Chern number $\pm 1$ ($\pm2$), while phase S is gapless.
The phase boundaries shown in Fig. \ref{fig:phase_diagrams}(b) are given
by four curves on the $(m,h_x)$ plane, 
\begin{align}
m&=\sqrt{1+h^2_x}+1, \label{mplus} \\
m&=\sqrt{1+h^2_x}-1,  \label{mminus}\\
h_x&=\sqrt{(m-1)^2+1}, \label{hminus}\\
h_x&=\sqrt{(m+1)^2+1}. \label{hplus}
\end{align}
They mark the closing of the gap in the continuum band.

{\bf Theorem 3}. In the thermodynamic limit $L\rightarrow \infty$, one of the edge modes of $H_2$ has dispersion
 \begin{equation}
E^+_{\text{edge}}(k_y)=+\sin k_y , \label{edgeE}
\end{equation}
with decay factor [defined in Eq. \eqref{def-lambda} below]
\begin{equation}
\lambda_+ = m-\cos k_y -h_x. \label{edgeBeta}
\end{equation}
In slab geometry, it is localized on the left (right) edge if $|\lambda_+|<1$ ($>1$).

{\bf Theorem 4}. The other edge mode has dispersion 
 \begin{equation}
E^-_{\text{edge}}(k_y)=-\sin k_y , \label{edgeEminus}
\end{equation}
with decay factor 
\begin{equation}
\lambda_- = m-\cos k_y +h_x \label{edgeBetaminus}.
\end{equation}
It is localized on the right (left) edge if $|\lambda_-|<1$ ($>1$). 

{\bf Theorem 5}. Phase C$_1$ is further partitioned into three regions (RR, LR, LL)
based on the localization of the two edge modes near $k_y=0$.
For example, in the LR region, the $E^+$ mode is localized on the left (L) edge, while the 
$E^-$ mode is localized on the right (R) edge. These three regions are separated 
by two lines:
\begin{align}
m&=h_x,\\
m+h_x&=2.
\end{align}
These lines do not correspond to gap closing. Rather, they mark the
divergence of the localization length, i.e., $|\lambda_{\pm}|=1$, at $k_y=0$.

{\bf Theorem 6}. In phase C$_2$, there are four edge modes at zero energy. Among them,
$E^{\pm}(k_y=\pi)$ are localized on the left edge, while $E^{\pm}(k_y=0)$ are localized 
on the right edge. 

{\bf Theorem 7}. Phase S is gapless with two edge modes localized on the left edge and
crossing $E=0$ at $k_y=\pi$. 

{\bf Theorem 8}. The energy eigenvalues of $\mathcal{H}_2$ are real for $m+\cos k_y> h_x$. 
For example, phase T at the bottom right corner of Fig. \ref{fig:phase_diagrams}(b) has {a} real spectrum.

Taken together, these eight theorems provide a clear characterization of the phases and the edge modes of model 2. These
analytical results agree with the numerical diagonalization of $H_2$ for large $L$ in slab geometry found in Ref. \cite{PhysRevB.98.165148}.
Below, we prove these theorems, and present a more detailed discussion of the phase diagrams, edge modes, and
topological invariants. 

\subsection{Continuum bands}

Theorem 1 has been proved back in Sec. II.B. Since the GBZ is a circle, $\beta$ $\in$ GBZ
can be parametrized by introducing a wave vector $\tilde{k}_x$:
 \begin{equation}
\beta =r_2 e^{i\tilde{k}_x}, \;\;\;\; \tilde{k}_x\in [-\pi,\pi].
\end{equation}
Then the continuum bands of $\mathcal{H}_2(\beta,k_y)$ can be found from Eq. \eqref{H2beta}.
After a little algebra, we find 
\begin{eqnarray}
E_c^2 (\tilde{k}_{x},k_{y})&=& 1+ m_y^2 -h_x^2 +(h_xr_- - m_yr_+)\cos \tilde{k}_x  \nonumber\\
&+&\sin^{2}k_{y}+i (h_xr_+ - m_yr_-)\sin \tilde{k}_{x}, \label{cont-band}
\end{eqnarray}
where the shorthand notation
\begin{eqnarray}
m_y&=&m-\cos k_y, \\
r_{\pm}&=& r_2 \pm r_2^{-1}.
\end{eqnarray}
In general, the eigenenergy $E_c$ is complex according to Eq. \eqref{cont-band}.  Within the region $m_y>h_x$, however, 
$h_xr_+ - m_yr_-=0$ and therefore $E^2_c$ is real.  By direct calculation, one can further show $E^2_c>0$ which proves Theorem 8.  

\begin{figure}[htpb!]
    \includegraphics[width=0.47\textwidth]{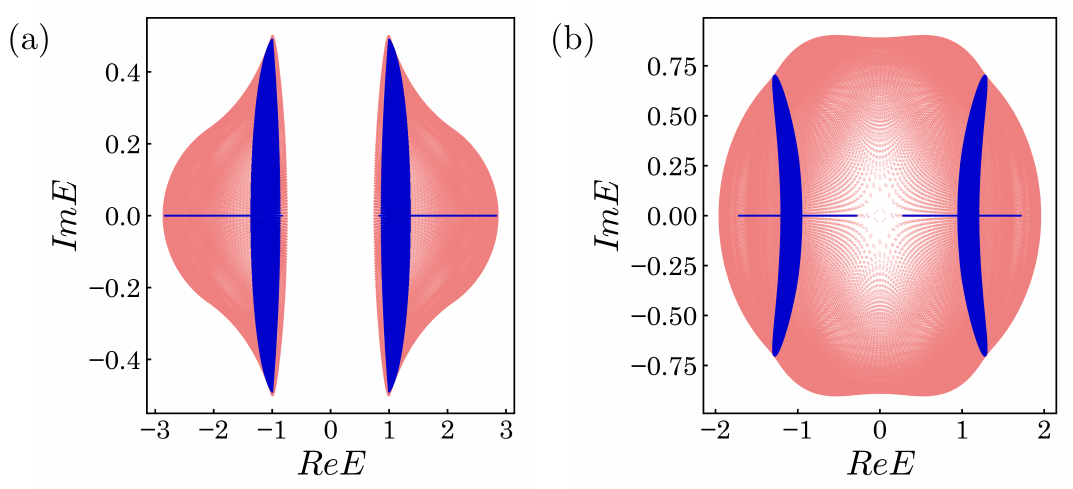}
    \caption{\label{fig:C2_Spectrum} Deviation of the continuum band spectrum (blue) from the 
    bulk spectrum (red) for model $H_2$. (a) Phase C$_1$ with $m=0.9$, $h_x=0.5$. (b) Phase C$_2$ with $m=0.15$, $h_x=0.9$. Notice the opening of the line gap in case (b) and the dramatic reconstruction of the band structure (red$\rightarrow$blue) as the boundary conditions change.
    }
\end{figure}

Inspecting the continuum band structure confirms that phase C$_1$ and C$_2$ have a line gap, while phase S is gapless.
Figure \ref{fig:C2_Spectrum} gives two examples of the continuum bands (in color blue) for phases C$_1$ and C$_2$, respectively.
It is illuminating to compare the continuum band $E^2_c$ above with the bulk spectrum:
of $H_2(k_x,k_y)$,
\begin{eqnarray}
E^2_{\text{bulk}}(k_{x},k_{y})&=&1+ m_y^2 -h_x^2 - 2m_y\cos k_x \nonumber\\
&+& \sin^{2}k_{y} + i2h_x \sin k_x.
\end{eqnarray}
This result clearly illustrates the highly nontrivial reconstruction of the band structure
in many non-Hermitian Chern insulators, $E^2_b\rightarrow E^2_c$, as 
the boundary conditions change from periodic to open along the $x$ direction. 
In Fig. \ref{fig:C2_Spectrum}, the bulk spectra (in color red) obviously deviate from the corresponding continuum bands (in blue). 
For example, in phase C$_2$ one would expect the system to be gapless from the bulk dispersion, but instead the continuum
band in the slab geometry develops a line gap, giving rise to a novel phase C$_2$.
Such band reconstruction is responsible for much of the rich behaviors of non-Hermitian Chern insulators
in the slab geometry.

Let us find out when the gap closes from the expression of $E^2_c(\tilde{k}_{x},k_{y})$. First, consider $k_y=0$, 
so $m_y= m-1$. For the case of $m_y>h_x$, let $z=\sqrt{m_y^2-h^2_x}$, then 
\begin{equation}
E_c^2 = 1+ z^2 -2z\cos \tilde{k}_x . 
\end{equation}
Obviously, $E_c^2=0$ requires $\cos \tilde{k}_x=1$ so the solution is $z=1$, i.e., $m_y^2-h^2_x=1$ leading directly to Eq. \eqref{mplus}. 
For the opposite case $m_y<h_x$, the gap touches down at $\tilde{k}_{x}=0$ or $\pi$ with
\begin{equation}
E_c^2 = (m_y\pm r_+/2)^2 - (h_x\pm r_-/2)^2.
\end{equation}
Thus $E_c^2=0$ leads to, after a little algebra, $h^2_x= m_y^2+1$ which gives Eq. \eqref{hminus}.
Similarly, the gap may close at $k_y=\pi$ with $m_y=m+1$ instead. Running the calculation again for $m_y>h_x$,
we are led to Eq. \eqref{mminus}, while for $m_y<h_x$, the result is Eq. \eqref{hplus}. Now we have found
all the phase boundaries summarized in Theorem 2.

\subsection{Edge modes}

Theorems 3 and 4 are not new results. 
The edge dispersions Eqs. \eqref{edgeE} and \eqref{edgeEminus} were established previously in Ref. \cite{PhysRevB.98.165148}.
For completeness, we briefly recount
the derivation here. This serves three purposes. First, it clarifies the origin of the analytical expression for $\lambda_\pm$ 
which we will use to establish Theorems 5-7. 
Second, we will apply the same approach to model 3 in the next section, where the calculation becomes more
challenging. Third, we find it fascinating that sinusoidal edge dispersion emerges not only for models
$H_2$ and $H_3$ (see Sec. V.B) but also for some driven quantum Hall systems \cite{PhysRevB.94.245128}. Thus, it is worthwhile to review
the main arguments.

Consider a semi-infinite system ($x\geq 0$) with an open boundary at $x=0$ (the left edge) described by the matrix $\mathcal{T}$ in 
Eq. \eqref{Tmatrix} with $L\rightarrow \infty$. Seeking a solution for the edge state, we try the ansatz
\begin{equation}
\psi = (\phi, \lambda \phi, \lambda^2 \phi, ...)^{T} ,  \label{def-lambda}
\end{equation}
where $\lambda$ is referred to as the decay factor, $\phi$ is a two-component spinor and $(...)^T$ means transpose. In terms of the $2\times 2$ matrices $A,B,C$ defined in Eqs. \eqref{Adef} to  \eqref{Cdef}, the eigenvalue
problem $\mathcal{T}\psi = E\psi$ reduces to
\begin{eqnarray}
\left[ A + B\lambda  \right] \phi &=& E \phi, \label{ABv}\\
\left[ C\lambda^{-1} + A+ B\lambda  \right] \phi &=& E \phi.
\end{eqnarray}
Following Ref. \cite{PhysRevB.98.165148}, we conclude that $C\phi=0$, and $\phi = (1,i)^{T}/\sqrt{2}$.
Plugging $\phi$ back into Eq. \eqref{ABv}, we are facing the following dilemma:
\begin{equation}
(m -\cos k_y-\lambda-h_x) 
\phi^* =
(E-\sin k_y) \phi .
\end{equation}
The only way for this equation to hold is for the two coefficients in the parenthesis to vanish. This proves 
Eqs. \eqref{edgeE} and \eqref{edgeBeta}. 

Note the energy of the edge state is always real, and crosses zero at $k_y=0$ or $\pi$.
Let us examine the spatial profile of this edge mode near $k_y=0$.  
It is localized on the left edge if the wave function decays with increasing $x$, that is, if
$|\lambda|<1$. According to Eq. \eqref{edgeBeta}, this occurs within the strip $h_x<m<2+h_x$ 
(recall we only focus on the first quadrant $m,h_x>0$). Outside this region on the $(m,h_x)$ plane, 
$|\lambda|>1$ so the solution $E^+$ describes a mode that grows with $x$, i.e.,  localized on the right edge.
(In the slab geometry, an edge mode on the right
edge still needs to satisfy the open boundary condition at $x=0$.)

The other edge mode solution can be worked out analogously by considering a semi-infinite
system occupying $x\leq 0$, with an open boundary on the right edge $x=0$. In this case, we seek solution of the type
\begin{equation}
\psi = (..., \lambda^2 \phi, \lambda \phi, \phi)^{T} , 
\end{equation}
with
\begin{eqnarray}
\left[ C\lambda + A   \right] \phi &=& E \phi,\\
\left[ C\lambda + A+ B\lambda^{-1}  \right] \phi &=& E \phi.
\end{eqnarray}
By repeating the argument in the preceding paragraph, it is straightforward to show Eqs. \eqref{edgeEminus} and \eqref{edgeBetaminus}. 
At $k_y=0$, we find that when $m+h_x<2$, $|\lambda|<1$, i.e., the edge state is localized on the right edge.
Otherwise, the solution represents a state on the left edge. Note that in the discussion above,
we have implicitly assumed the continuum band structure has a gap at $k_y=0$. Otherwise, the solution
does not qualify as an edge state.

\begin{figure}[htpb!]
    \centering
    \includegraphics[scale=1]{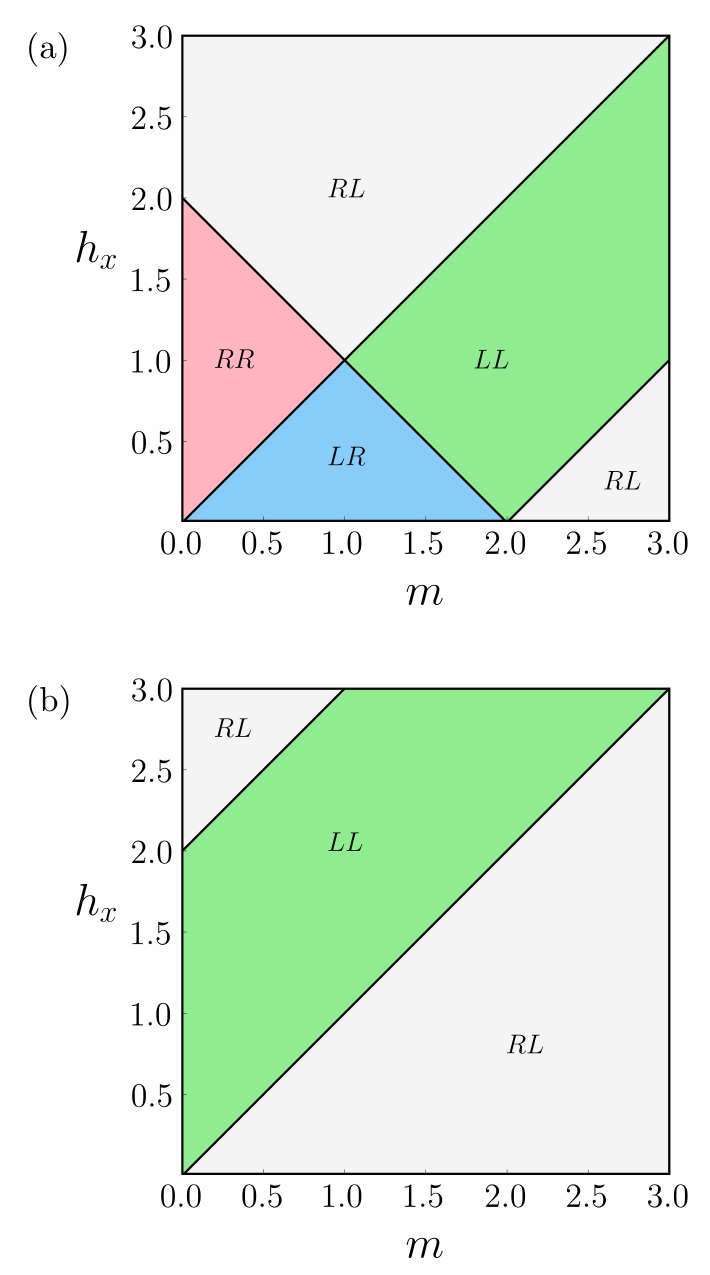}
    \caption{\label{fig:H2_loalization} The localization phase diagrams for the edge modes of model $H_2$ near (a) $k_y=0$ and (b) $k_y=\pi$. The first (second) capital letter indicates the localization of the $E^+$ ($E^-$) edge mode. Combining these
    results with the phase boundary in Theorem 2 fixes the phase diagram of model $H_2$ in Fig. \ref{fig:phase_diagrams}(b).
    See main text for details.}
\end{figure}

These results regarding the location of the edge modes near $k_y=0$ can be combined to yield the ``localization phase diagram"
shown in Fig. \ref{fig:H2_loalization}(a). We can identify four regions on the $(m,k_y)$ plane: LL, RR, LR, and RL. Here the first capital letter
indicates whether the $E^+(k_y\sim 0)$ mode resides on the left (L) or right (R) edge, while the second letter 
describes the location of the $E^-(k_y\sim 0)$ mode. In particular, the RR, LR, and LL regions are separated by two 
lines, $m=h_x$ and $m+h_x=2$, where $|\lambda_\pm|=1$. This proves Theorem 5.

The edge states $E^{\pm}$ also cross zero energy at $k_y=\pi$ inside phase C$_2$ and phase S.
From the expression for $\lambda_{+}$ in Eq. \eqref{edgeBeta}, we conclude that $E^{+}(k_y\sim \pi)$ is localized 
on the left edge for $m<h_x<m+2$. Similarly, from $\lambda_{-}$, we find that $E^{-}(k_y\sim \pi)$ is always localized 
on the left edge in the first quadrant. The resulting  ``localization phase diagram" for the edge modes 
near $k_y=\pi$ is summarized in Fig. \ref{fig:H2_loalization}(b). If we overlay Figs. \ref{fig:H2_loalization}(a) and \ref{fig:H2_loalization}(b), we are led to Theorem 6: 
Inside phase C$_2$, the two modes $E^\pm(k_y\sim 0)$ are within the region of RR, while 
$E^\pm(k_y\sim \pi)$ are within the region of LL. In other words, out of the four edge modes crossing the zero energy,
two of them are on the left edge, and the other two are on the right edge. We stress once again that
such a scenario is only possible in non-Hermitian Chern insulators. 

\subsection{Chern number}

To characterize all gapped phases of model $H_2$, we compute the Chern numbers. 
The starting point is the analytically continued, non-Hermitian Hamiltonian $\mathcal{H}_2(\beta,k_y)$ in Eq. \eqref{H2beta} with $\beta=r_2(k_y)e^{i\tilde{k}_x} \in \mathrm{GBZ}$. 
The left and right eigenstates of $\mathcal{H}_2$ are defined as
\begin{eqnarray} 
 \mathcal{H}_2\ket{\psi_{\ell}}=E_\ell\ket{\psi_{\ell}}, \label{eigenpsi} \\
\bra{\xi_\ell} \mathcal{H}_2= \bra{\xi_{\ell}} E_\ell.
 \end{eqnarray}
Here $\ell=\pm$ is the band index, and the dependence on $(\beta,k_y)$ is suppressed.
The right eigenstates $\left\{\ket{\psi_{\ell}}\right\}$ are linearly independent but not necessarily orthogonal \cite{Brody_2013}. Instead, 
we require them to satisfy the biorthogonal normalization condition, $\braket{\xi_{\ell}|\psi_{\ell'}}=\delta_{\ell,\ell'}$. The (generalized) Chern number is defined as
an integral of the Berry curvature over the GBZ surface,
\begin{eqnarray}
C_\ell=\frac{1}{2\pi i}\int_{\mathrm{GBZ}_s}d\tilde{k}_{x} dk_y\epsilon^{ij}\partial_{i}\bra{\xi_\ell}\partial_{j}\ket{\psi_\ell}.\label{chern}
\end{eqnarray}
Here $i,j=\tilde{k}_x,k_y$ are the two independent directions on the GBZ surface, with repeated indices summed over.

According to Eq. \eqref{radiusr2}, the GBZ curve as a circle shrinks to a dot ($r_2=0$) when
$\cos k_y = -(m+h_x)$
and the radius $r_2$ diverges when
$\cos k_y = -(m-h_x)$. 
Thus, rigorously speaking, the Berry curvature becomes ill defined at these singular points of $k_y$. 
To yield a sensible result, the integral in Eq. \eqref{chern} must be understood as the principal value.
An efficient, gauge-invariant way to numerically evaluate the Chern number is to partition the BZ into little patches
and find the flux through each patch, e.g., by taking the trace of the Berry connection along the boundary of the patch \cite{doi:10.1143/JPSJ.74.1674}. 
This algorithm can be generalized to compute the Chern number over the GBZ$_s$, as long as one carefully avoids hitting 
the singular points along the patch boundaries. To understand why this procedure works, imagine continuously deforming 
the GBZ$_s$ only at the vicinity of these singularities so it becomes closed and smooth, leaving 
the patch boundary intact. Thanks to Gauss's theorem, the total flux stays the 
same during the deformation, as long as the small deformation does not encounter any magnetic charge. Then, the Chern 
number calculated on the deformed smooth
GBZ$_s$ is well-defined, and has the same value as the original GBZ$_s$ with integrable singularities.
We find the resulting Chern number for phase C$_1$ (C$_2$) is $\pm 1$ ($\pm 2$), which 
completes the proof of Theorem 2.

\begin{figure}[htpb!]
    \centering
    \includegraphics[scale=1]{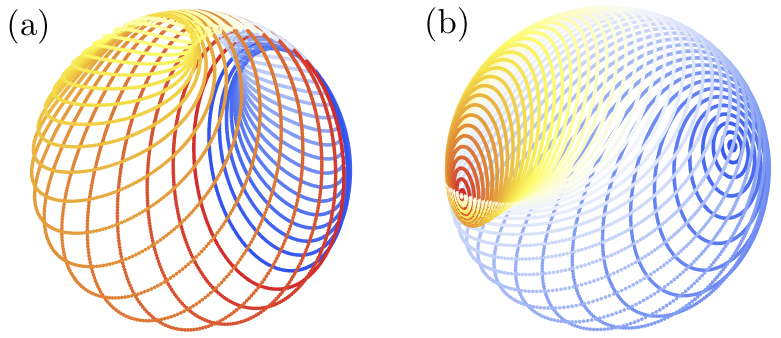}
    \caption{\label{fig:H2_chern} Counting the Chern number by the wrapping of eigenvectors of model $H_2$ on the 
    Bloch sphere. (a) Phase C$_1$ with $m=1.5$, $h_x=0.3$. Different colors correspond to a discrete set of $k_y$ values from $0$ to $2\pi$, while the data points of the same color depict varying $\tilde{k}_x$ for a given $k_y$. As $(\tilde{k}_x,k_y)$ transverse the entire GBZ$_s$, the eigenvector covers the Bloch sphere exactly once.  
     (b) Phase C$_2$ with $m=0$, $h_x=1$. Here only $k_y$ values restricted to $[0,\pi]$ are shown, and they are sufficient to cover the whole Bloch sphere. Thus, the Chern number is two. In the trivial phase T (not shown), the eigenvectors cannot cover the entire Bloch sphere.}
\end{figure}

To cross-check the numerical calculation of the Chern number, we adopt a complementary scheme to visualize and characterize
the topology of the gapped phases. The eigenvalue problem of $\mathcal{H}_2$ in Eq. \eqref{eigenpsi} defines 
a mapping from {the} GBZ$_s$ to the Bloch sphere once the eigenvector of $\mathcal{H}_2$ is parametrized using the polar angle $\theta$ and 
the azimuthal angle $\phi$,
\begin{equation}
(\beta,k_y) \in \mathrm{GBZ}_s \mapsto  |\psi\rangle = 
e^{i\chi}\left(
\begin{array}{c}
\cos\frac{\theta}{2}   \\  
\sin\frac{\theta}{2}e^{i\phi}
\end{array}
\right), \label{blochsphere}
\end{equation}
with $\chi$ the overall phase.  Then one can define the Chern number as the number of times the images 
of $(\beta,k_y)$ cover the whole Bloch spheres as it varies throughout the GBZ$_s$. Figure \ref{fig:H2_chern} shows the wrapping
for phases C$_1$ and C$_2$. They agree with the numerical integration results above. This approach circumvents
the subtleties regarding Berry curvature near the singular points of GBZ. Interestingly, it also provides a geometric
picture of these singularities. Direct analytical calculation reveals that, at these singular points, the eigenstates lie within the equator
of the Bloch sphere. More specifically, 
$r_2\rightarrow 0$ and $\infty$ corresponds to the eigenvector pointing along the $\mp y$ axis, respectively, i.e., $\theta=\pi/2$ and
$\phi=\mp \pi/2$. These two points are visible in Fig. \ref{fig:H2_chern} as the center of the small concentric red and blue rings.

\subsection{The gapless phase S}

\begin{figure}
    \centering
    \includegraphics[width=0.4\textwidth]{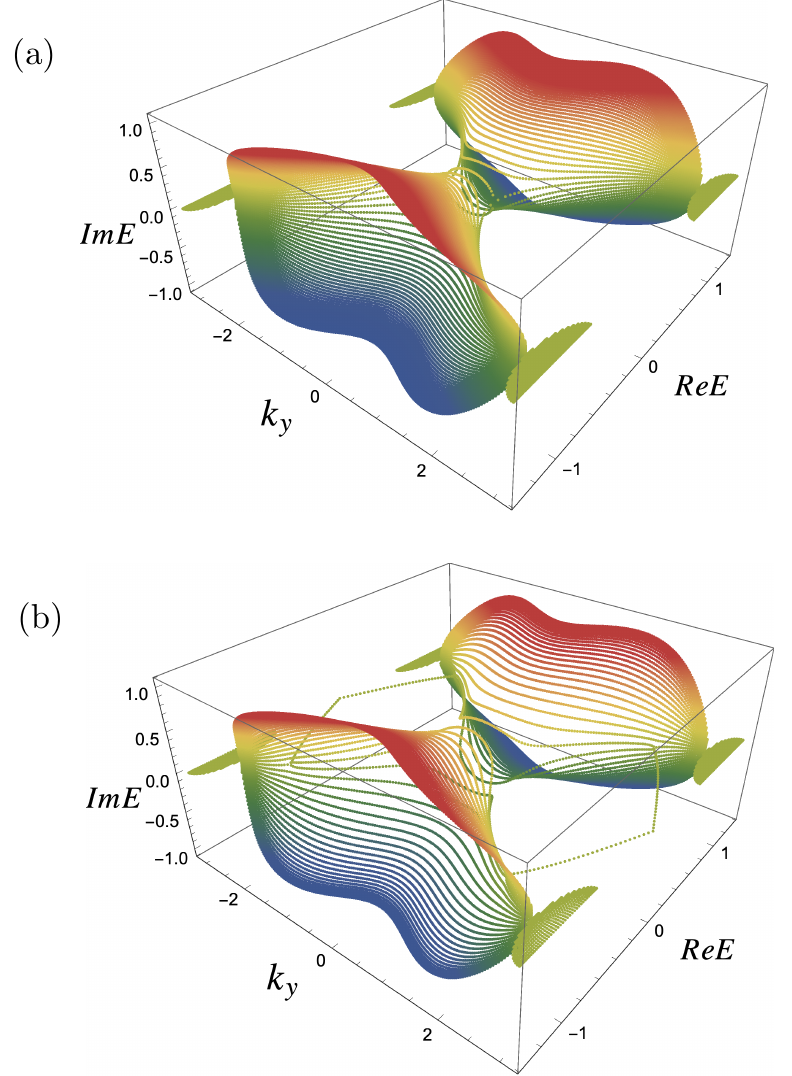}
    \caption{\label{fig:s_phase} The gapless phase S of model $H_2$. (a) The continuum band spectrum from analytical calculation. (b) Spectrum from numerical diagonalization of a slab with $L=50$, showing a pair of edge states crossing $E=0$ at $k_y=\pi$. Both are localized at the left edge. $m=0.3$, $h_x=1.25$.}
\end{figure}

Phase S is gapless, and it has no analog in Hermitian Chern insulators. When the spectrum (in slab geometry) is plotted on the complex energy plane, there is no line gap (but there is a point gap around $E=0$).  
The continuum band spectrum in Fig. \ref{fig:s_phase}(a) forms a single connected surface with ``holes" in the 
space of $(\mathrm{Re}E,\mathrm{Im}E,k_y)$. The finite-size spectrum in Fig. \ref{fig:s_phase}(b) clearly shows a pair of edge states crossing $E=0$ at $k_y=\pi$ 
as described by Eqs. \eqref{edgeE} and \eqref{edgeEminus}. From the localization phase diagram
of these modes in Fig. \ref{fig:H2_loalization}(b), it is clear that both modes localize on the left edge, which proves Theorem 7 and is confirmed by numerical results. Note the good agreement between Figs. \ref{fig:s_phase}(a) and \ref{fig:s_phase}(b) for $L=50$. For large $L$, the diagonalization of the non-Hermitian Hamiltonian is prone to numerical instabilities, and the resulting spectrum starts to show large fluctuations due to numerical error and then becomes unreliable. This further reinforces the usefulness of the GBZ and the analytical approach we advocate here which operates directly in the $L\rightarrow \infty$ limit.

In passing, we mention that in the phase diagram Fig. \ref{fig:phase_diagrams}(b), the trivial region T at the bottom-right corner has {a} line gap and real spectrum. In comparison, the spectrum of the T region at the left top corner is complex. The region GL is gapless, there are edge modes but they do not cross zero energy.
To summarize, analytical results obtained for $\mathcal{H}_2$ in this section capture all the main features 
of the phase diagram, edge states, and topological characterization of each gapped phase. This example 
illustrates the capability as well as the subtleties of this approach based on {the} GBZ. We will apply the approach to another in the next section.

\section{Model 3}

In this section, we analyze $H_3$, the third model of non-Hermitian Chern insulators defined in Eq. \eqref{H3}.
Compared to model $H_1$ and $H_2$ above, this model introduces a unique feature: the hopping between neighboring unit cells is non-reciprocal. More specifically, the intra-orbital hopping amplitudes to the left and right are given by $1+t_1$ and $1-t_1$ respectively. The finite $t_1$ term makes it more challenging to find the GBZ, the continuum band structure, the phase diagram, and the edge states.
But these tasks are still manageable thanks to the techniques developed in Secs. II and IV.

For the sake of clarity, we summarize our main results into Theorems 9-13 below. Hereafter, the term phase diagram refers to the phase diagram for model 3 in slab geometry (with open boundaries at $x=0,L$ and periodic boundary conditions along $y$) in the limit of $L\rightarrow \infty$, and we shall restrict our attention to the first quadrant of the parameter space, $m,t_1>0$.
It is straightforward to generalize the analysis to other parameter regions.

{\bf Theorem 9}. For $k_y=0$, the GBZ is a circle on the complex $\beta$ plane with radius $r_3$ given by Eq. \eqref{radius3}.

{\bf Theorem 10}. The phase diagram of $H_3$ is mirror symmetric with respect to $m=1$. It consists of six regions  shown in Fig. \ref{fig:phase_diagrams}(c). 
The continuum bands of phase T have a line gap and {are} topologically trivial (i.e., their Chern numbers are zero). Region C$_1^{\mathrm{LR}}$ and C$_1^{\mathrm{LL}}$ belong to the same gapped phase with band Chern numbers $\pm 1$.  Phase GL$_1$, GL$_2$, and S$'$ are gapless.

{\bf Theorem 11}. Region C$_1^{\mathrm{LR}}$ is bounded by two straight lines, $m=2$ and $t_1=1$. All other phase boundaries correspond to 
gap closing at $E=0$ and $k_y=0$, and are determined by solving an algebraic problem, Eqs. \eqref{quartic-eq-0} and \eqref{xi23-equal} below.
In particular, Phase C$_1^{\mathrm{LL}}$, GL$_1$, and S$'$ meet at the tricritical point $m=1$ and $t_1=\sqrt{2}$. 

{\bf Theorem 12}. One of the edge states has dispersion
 \begin{equation}
E_{\text{edge}}^+=\sin k_y.
\end{equation}
It is always localized the left edge for $t_1>0$.

{\bf Theorem 13}. The other edge state has dispersion
 \begin{equation}
E_{\text{edge}}^-=-\sin k_y.
\end{equation}
It is localized on the right (left) edge if $t_1<1$ ($t_1>1$).

\subsection{Phase boundaries}

\begin{figure}[htpb!]
    \centering
    \includegraphics[width=0.47\textwidth]{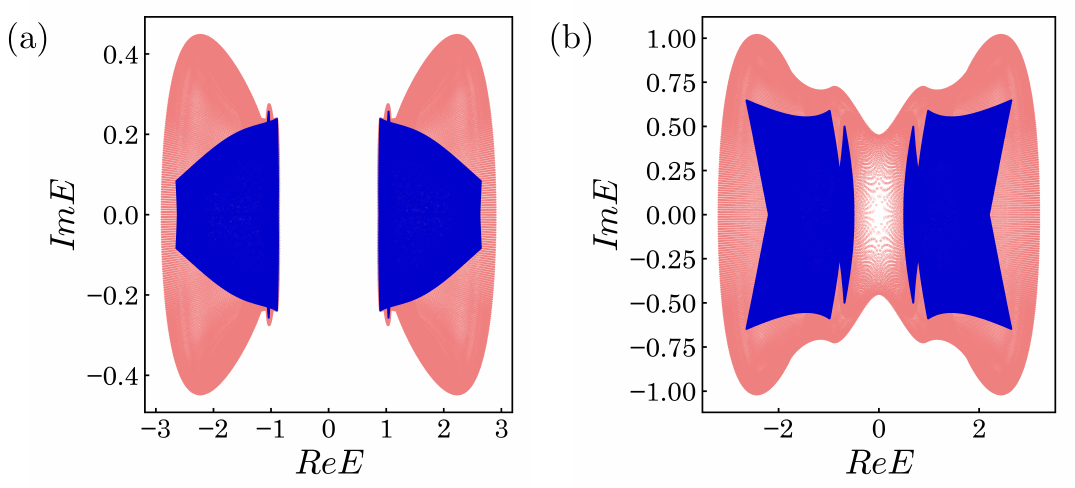}
    \caption{\label{fig:H3_spectra} Comparing the continuum band spectrum $E_c$ (in blue), defined for the slab geometry
    in the limit of $L\rightarrow \infty$,  with the bulk energy spectrum $E_{\text{bulk}}$ (in red) for model $H_3$. 
 (a) For phase C$_1^{\mathrm{LR}}$ with $m=0.9, t_1=0.5$. (b) For phase C$_1^{\mathrm{LL}}$ with $m=1.2$, $t_1=1.1$. Notice the gap opening as
 the boundary conditions change in case (b). 
    }
\end{figure}

The rough contour of the phase boundaries can be obtained from numerical diagonalization of the matrix $\mathcal{T}$ for finite $L$. For example, one can monitor
the minimum magnitude of the eigenenergy, $|E|_{\text{min}}$. On the one hand, this quantity is finite for gapped phases (e.g., phase T) that are topologically trivial, i.e., have
no edge states, or gapless phases (e.g., phase GL$_2$) that avoid $E=0$. On the other hand, it vanishes for topological phases with edge modes (e.g., phase C$_1$ and S$'$). Such a quick scan, however, has trouble in locating the precise boundary of phase GL$_1$ which is gapless and contains $E=0$. In particular, one
notices that the numerical spectrum depends sensitively on $L$.  It is well-known that diagonalization of large non-Hermitian matrices
can experience numerical instabilities, and caution must be exercised before trusting their accuracies, see Refs. \cite{PhysRevLett.125.226402,PhysRevLett.122.250201,RevModPhys.93.015005,alma9946931517304105,REICHEL1992153} for detailed discussions. 
Thus, an alternative, algebraic method that works well in the limit of $L\rightarrow \infty$ is desired.

We reiterate the key point that for non-Hermitian Chern insulators, the knowledge of the bulk spectrum may offer little help 
in determining or understanding its slab phase diagram as we have witnessed in the case of $H_2$. For model $H_3$ here, the bulk energy $E_{\text{bulk}}(k_x,k_y)$
closes its gap along the line of $m=2$ at $k_x=k_y=0$, and along the line of $t_1=1$ at $k_y=0$ and $\cos k_x=m-1$. While
the $m=2$ line agrees with the phase boundary between C$_1^{\mathrm{LR}}$ and T, the $t_1$ line, as we shall show below, is not a phase transition line.
Moreover, the bulk spectrum fails to predict phase C$_1^{\mathrm{LL}}$, GL$_1$, and S$'$.

Now we show that all the phase boundaries in the limit of $L\rightarrow \infty$ can be worked out from the
analytically continued Hamiltonian $\mathcal{H}_3(\beta \in \mathrm{GBZ}, k_y)$. For any given value of $k_y$, the GBZ can be 
computed using the two algorithms outlined in Secs. II.C and II.D. 
As a simple example, let us consider the cutline $m=1$ with varying $t_1$, and focus on $k_y=0$. In this case, Theorem 9 was already proved back in Eqs. (39) and (40). 
With {the} GBZ being a circle, we can parametrize it using a fake wave vector $\tilde{k}_x$:
 \begin{equation}
\beta = r_3 e^{i\tilde{k}_x}.
\end{equation}
Then the continuum band spectrum simplifies for $t_1<2$:
 \begin{equation}
E^2(\tilde{k}_x,k_y=0) = 1-\frac{{t_1}^2}{2} -it_1\sqrt{4-t_1^2}\sin (2 \tilde{k}_x).
\end{equation}
Immediately, we see that for $\tilde{k}_x=0$, $E$ vanishes when $t_1=\sqrt{2}$,
which marks the tricritical point between phases C$_1^{\mathrm{LL}}$, GL$_1$, and S$'$.
Away from the central $m=1$ line, {an} analytical solution seems out of reach, and the GBZ has to be found numerically 
to yield the continuum bands.

For the purpose of finding the phase boundaries, however, it is not necessary to gain a full knowledge of either the GBZ or the continuum bands.
It turns out that for model 3, one only needs to check when the energy gap closes at $E=0$ at some $k_y$ values, say $k_y=0$. 
Below, we outline how this problem can be reduced to solving a quartic equation. Following the notation introduced in Sec. II,
let  $\xi_i$ with $i=1,2,3,4$ be the four solutions to the
quartic equation
\begin{equation}
\epsilon(\beta) = a\beta^2 + b/\beta^{2}+c\beta+d/\beta+f =0, \label{quartic-eq-0}
\end{equation}
with their magnitudes ordered according to Eq. \eqref{orderxi}. In other words, $\xi_i$ are the preimages of $\epsilon=0$.
Then a gap closing, $E(\beta\in \mathrm{GBZ},k_y=0)=0$, requires the two roots 
of the quartic Eq. \eqref{quartic-eq-0} to have the same amplitude:
\begin{equation}
|\xi_2|= |\xi_3|. \label{xi23-equal}
\end{equation}
Recall that the coefficients $a$ to $f$ depend on parameter $m$ and $t_1$.
Thus, to find points on the $(m,t_1)$ plane where Eq. \eqref{xi23-equal} is satisfied, we can simply follow a given horizontal or vertical
cut and plot $|\xi_i|$ to see where $|\xi_2|$ and $|\xi_3|$ intersect. (While the roots of quartic equations are analytically
known, they are unwieldy to manipulate so we opt to find and compare $|\xi_2|$ and $|\xi_3|$ numerically.)
The phase boundaries obtained this way are summarized in Fig. \ref{fig:phase_diagrams}(c). They agree with the rough outline from numerical diagonalization of finite-size
slabs. The main advantage of the algebraic approach is that the phase boundaries (e.g., that of phase GL$_1$) can be obtained precisely. Compared to model 2, here the phase boundaries 
of model 3 are no longer simple analytical curves, but we still find it remarkable that it follows from the solution of a quartic equation.

Once these boundaries are drawn from the gap closing condition, we can investigate the continuum bands in each region.
For example, one can confirm that C$_1$ and T are gapped with {a} line gap, while GL$_1$, GL$_2$, and S$'$ are gapless.
Figure \ref{fig:H3_spectra} highlights the contrast between the bulk spectrum (red) and the slab spectrum (blue) in the limit of $L\rightarrow \infty$ obtained from $\mathcal{H}_3$. For example, the existence of the line gap (and the edge states) within phase 
C$_1^{\mathrm{LL}}$ would have been completely missed by only considering $E_{\text{bulk}}(k_x,k_y)$.
To unambiguously identify each phase, in the next subsection we proceed to look into their edge spectra and topological invariants.
For example, we shall see that regions C$_1^{\mathrm{LR}}$ and C$_1^{\mathrm{LL}}$ are divided by a transition line at $t_1=1$ where
the edge modes change location.

\subsection{The dispersion and location of edge modes}

\begin{figure}[htpb!]
    \centering
    \includegraphics[scale=1]{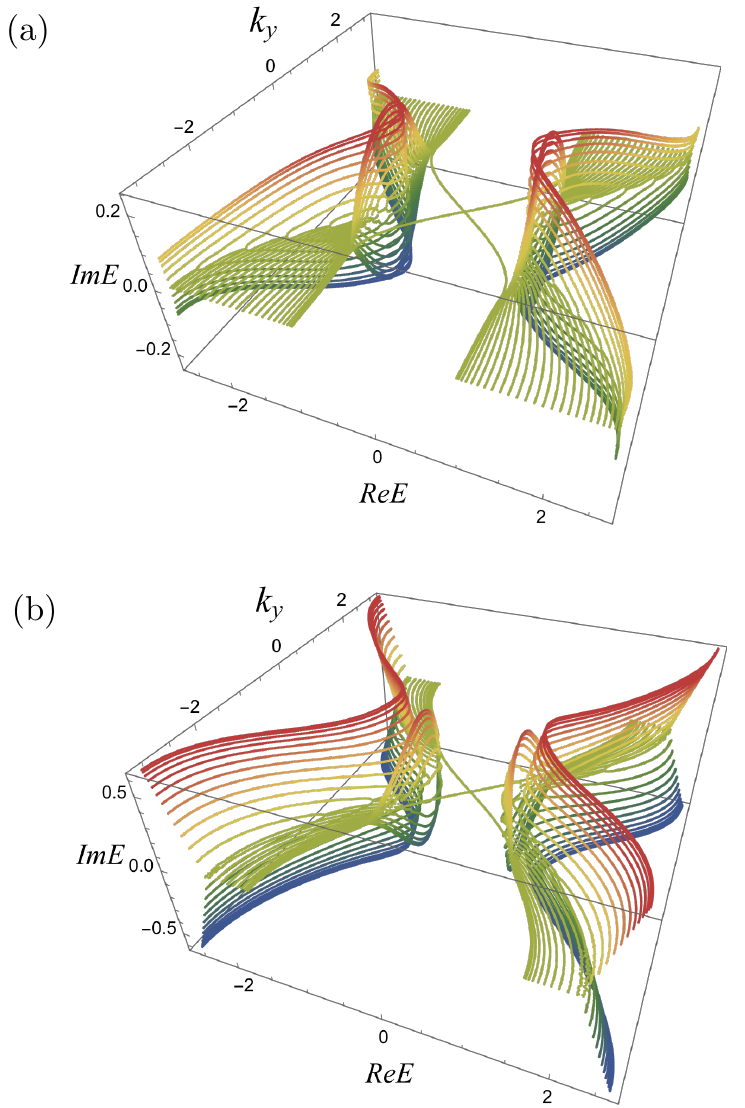}
    \caption{\label{fig:H3_energy}{The spectrum of $H_3$ in slab geometry (with slab width $L=35$) inside (a) the C$_1^\mathrm{LR}$ region for $m=0.9$, $t_1=0.5$, and (b) the C$_1^\mathrm{LL}$ region with parameters $m=1.2$ and $t_1=1.1$. In both cases, a pair of edge modes $E^\pm_{\text{edge}}=\pm \sin k_y$ transverse the line gap.  Note, however, that for case (a), one mode is localized on the right edge and the other on the left edge, whereas in (b), both edge modes are localized on the right edge. The transition in the localization behavior occurs at line $t_1=1$ which separates region C$_1^\mathrm{LR}$ and C$_1^\mathrm{LL}$.}}
\end{figure}

To find the edge states and prove Theorems 12-14 for model $H_3$, we once again face the big matrix 
$\mathcal{T}$ in Eq. \eqref{Tmatrix}. But this time its submatrices are given by 
\begin{eqnarray}
 A &=& [m-\cos k_y] \sigma_z +\sin k_y\sigma_y,\\
B &= &[(-t_1-1)\sigma_z + i\sigma_x]/2,\\
C &=& [(t_1-1)\sigma_z - i\sigma_x]/2.
\end{eqnarray}
The overall strategy is the same as in Sec. IV.B. The wave functions of the edge modes, however, become
more cumbersome due to the non-reciprocal hopping $t_1$. 

First, consider the semi-infinite geometry ($x\geq 0$) with an open boundary at $x=0$, the left edge. 
To solve the eigenvalue problem $\mathcal{T}\psi = E\psi$, let us write $\psi$ as 
\begin{equation}
\psi = (v_1, v_2, v_3, ...)^{T} , \label{leftrecipe}
\end{equation}
where $v_i$ is a two-component spinor. This leads to
\begin{eqnarray}
&A v_1 + B v_2 = E v_1, \label{vAB} \\
&Cv_{n-1} + Av_n+ B v_{n+1}  = E v_n, \;\;\; (n\geq 2). \label{morev}
\end{eqnarray}
In the limit $t_1=0$, we conclude $v_1=(1,i)^T/\sqrt{2}$ and $E=\sin k_y$, which
we take as the guess solution for the general case.  From $v_1$, all other $v_n$ can be found inductively using Eqs. \eqref{vAB} and \eqref{morev}.
Exploiting the properties of Pauli matrices, after some algebra we conclude that
\begin{equation}
v_n= \lambda_n v_1, 
\end{equation}
where $\lambda_n$ is a number. Within this ansatz, Eqs. \eqref{vAB} and \eqref{morev} become the following recursion relation for $\lambda_n$:
\begin{equation}
(2+t_1) \lambda_n= 2(m-\cos k_y) \lambda_{n-1}  + t_1 \lambda_{n-2}, 
\end{equation}
with the initial condition
\begin{equation}
\lambda_0 = 0,\;\; \lambda_1=1. \label{boundary-lambda}
\end{equation}
We seek a solution of the power-law form $\lambda_n=\lambda^n$.
Here $\lambda$ describes the decay (or growth) of the wave function $\{v_n=\lambda_n v_1\}$, and must obey the quadratic equation
\begin{equation}
(2+t_1)\lambda^2 - 2(m-\cos k_y) \lambda - t_1= 0. \label{quadratic-lambda}
\end{equation}
This equation has two solutions which we call $\lambda_\pm$. The
general solution is then the superposition $\lambda_n=c_1\lambda_+^n + c_2\lambda_-^n$. The initial condition Eq. \eqref{boundary-lambda}
fixes the coefficients $c_{1,2}$. The final result is 
\begin{equation}
\lambda_n=  \frac{\lambda_+^n - \lambda_-^n}{\lambda_+ - \lambda_-}=\sum_{j=0}^{n-1} \lambda_+^{n-1-j} \lambda_-^j.
\end{equation}
It decays with increasing $n$ if and only if $|\lambda_\pm|<1$, or equivalently, $|\lambda_+\lambda_-|<1$. 
This condition can be further simplified by recalling
Vieta's formula,
\begin{equation}
|\lambda_+\lambda_-| = \frac{t_1}{2+t_1}<1 \label{blowup}
\end{equation}
for $t_1>0$. Note that this criterion is independent of $m$ or $k_y$.
It follows that that the edge state with energy $E^+_{\text{edge}}=\sin k_y$ 
is always localized on the left edge for $t_1>0$. This proves Theorem 12.

The calculation of the other edge mode proceeds similarly. For a semi-infinite
system ($x\leq 0$) with an open boundary at $x=0$, let us label the wave function as
\begin{equation}
\psi = (..., u_{3}, u_{2}, u_{1})^{T}. \label{rightrecipe}
\end{equation}
With ansatz $E^-_{\text{edge}}=-\sin k_y$, $u^T_1=(i,1)/\sqrt{2}$, and $u_n=\lambda^n u_1$, one finds
that the decay factor $\lambda$ is determined by
\begin{equation}
(2-t_1)\lambda^2 - 2(m-\cos k_y) \lambda + t_1= 0. \label{quadratic-lambda-minus}
\end{equation}
This result can also be obtained from Eq. \eqref{quadratic-lambda} by symmetry arguments and replacing $t_1\rightarrow -t_1$.
For $t_1>0$, the magnitudes of the two solutions satisfy
\begin{equation}
|\lambda_+\lambda_-| = \frac{t_1}{|2-t_1|}. \label{magcomb}
\end{equation}
Thus, the $E^-$ edge mode is localized on the right edge if $t_1<1$, and on the left edge if 
$t_1>1$,  proving Theorem 13. The transition occurs at $t_1=1$.

It is worthwhile to take a closer look at the $E^-$ solution above in the region $t_1>1$.
At $t_1=2$, the matrix $C$ becomes singular with a vanishing determinant and its inverse becomes ill defined.
Accordingly, $|\lambda_+\lambda_-| $ diverges according to Eq. \eqref{magcomb}. We emphasize that
there is nothing physically singular at this point. To get a clearer picture, we must recognize that once 
$t_1$ exceeds 1 and the $E^-$ mode is localized on the left edge, it is much more natural to find its wave function
by starting from the left boundary, rather than from the right boundary as done in Eq. \eqref{rightrecipe}. 
More explicitly, we repeat the same recipe as prescribed in Eq. \eqref{leftrecipe}, but this time with ansatz 
$E^-_{\text{edge}}=-\sin k_y$ and $v^T_1=(i,1)/\sqrt{2}$ instead. The corresponding decay factor now satisfies the equation
\begin{equation}
t_1\lambda^2 - 2(m-\cos k_y) \lambda - (t_1-2)= 0. \label{thirdlambda}
\end{equation}
The magnitudes of its two solutions obey
\begin{equation}
|\lambda_+\lambda_-| = \frac{|t_1-2|}{t_1}.
\end{equation}
Compared to Eqs. \eqref{magcomb}, 
here the roles of $t_1$ and $(t_1-2)$ are switched, now that we seek the edge state wave function starting from the left boundary. 
It follows that the $E^-_{\text{edge}}$ mode is localized on the left edge if $t_1>1$, and on the right edge if $t_1<1$. 
This is consistent with our result obtained in the preceding paragraph and provides an alternative proof of Theorem 13.
The calculation here also yields the decay factor along the line $t_1=2$, where $Cv_1=0$. In this case, 
Eq. \eqref{thirdlambda} reduces to a linear equation, and we have $\lambda =2(m-\cos k_y)/t_1$, and $\lambda_n=\lambda^n$.

To summarize, within the region C$_1^{\mathrm{LR}}$, the two modes $E^{\pm}_{\text{edge}}$ reside on opposite edges of the slab. In region C$_1^{\mathrm{LL}}$ and phase S$'$, they both reside on the left edge, which is impossible for Hermitian Chern insulators. 
At the transition line $t_1=1$, where
$|\lambda_+\lambda_-|=1$, the $E^{-}_{\text{edge}}$ mode permeates into the bulk, and therefore strictly speaking is no longer an ``edge mode."  
These analytical results agree with the edge states obtained from numerical diagonalization of finite systems, see Fig. \ref{fig:H3_energy}.

\subsection{Chern numbers}

For each given value of $k_y$, the GBZ curve of $\mathcal{H}_3(\beta,k_y)$ can be found by following the recipes described in Section II. 
As $k_y$ is varied from $-\pi$ to $\pi$, the GBZ curve deforms to produce a 2D surface GBZ$_s$ defined in Eq. \eqref{GBZS}, which is continuous
but may have singularities. Figure \ref{fig:H3_GBZ} shows an example of GBZ$_s$ for $m=0.6$ and $t_1=0.4$.
Once the GBZ surface is known, one can proceed to compute the Chern numbers using Eq. \eqref{chern} by discretizing the GBZ$_s$
into patches. One can verify that region C$_{1}^{\mathrm{LR}}$ and C$_{1}^{\mathrm{LL}}$ have the same Chern numbers $\pm 1$, while region T is topologically trivial with Chern number zero.

\begin{figure}[htpb!]
    \centering
    \includegraphics[scale=1]{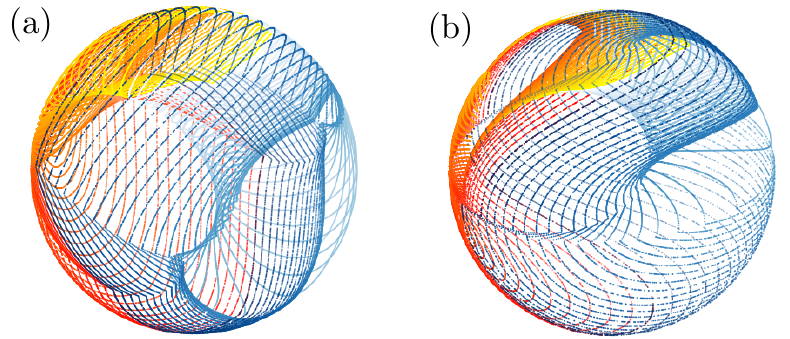}
    \caption{\label{fig:H3_chern} {Chern number determined from the eigenvectors of model $H_3$ on the 
    Bloch sphere. (a) Region C$_1^\mathrm{LR}$ with $m=0.6$, $t_1=0.4$. Different colors correspond to a discrete set of $k_y$ values from $0$ to $2\pi$, while the data points of the same color depict transversing the GBZ curve for a given $k_y$. As $(\beta,k_y)$ varies throughout the GBZ$_s$, the eigenvector covers the Bloch sphere exactly once. (b) Region C$_1^\mathrm{LL}$ with $m=1.2$, $t_1=1.1$. Here, again $k_y$ varies from $0$ to $2\pi$ and the eigenvector covers the Bloch sphere exactly once as $(\beta,k_y)$ transverse the entire GBZ$_s$. In the trivial phase T (not shown), the eigenvectors do not cover the whole Bloch sphere.
    }}
\end{figure}

Since the discretization of GBZ$_s$ is numerically involved, we also extract the Chern number by counting
the number of times the right eigenvector in Eq. \eqref{blochsphere} wraps around the Bloch sphere, as done for model 2 in Sec. IV.C.
According to Fig. \ref{fig:H3_chern}, within phase C$_1$, including region C$_1^\mathrm{LR}$ shown in panel (a) and region C$_{1}^{\mathrm{LL}}$ shown in panel (b), the eigenstate wraps the Bloch sphere once. Within the trivial phase T,
the eigenstate does not cover the entire sphere. This verifies that C$_{1}^{\mathrm{LR}}$ and C$_{1}^{\mathrm{LL}}$ differ topologically from phase T, and finishes the proof of Theorem 10. We emphasize that Theorem 10 here relies on numerical 
evaluation of the Chern number and the phase boundary from a well-defined algebraic problem. This is slightly different from the proof of, e.g., Theorem 7,
where the phase boundary is simple and analytically known.

\subsection{The gapless phase S$'$}

\begin{figure}[htpb!]
    \centering
    \includegraphics[width=0.4\textwidth]{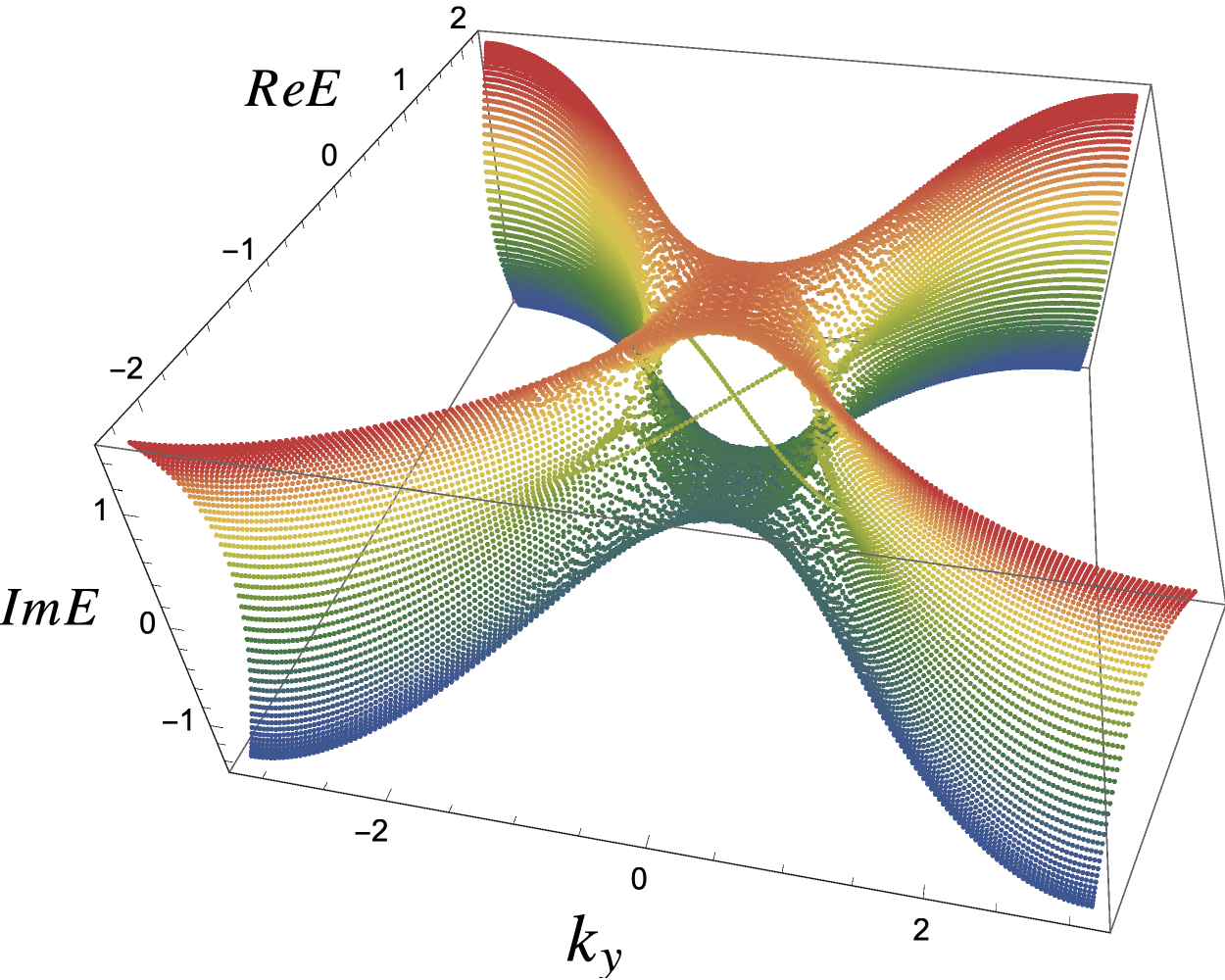}
    \caption{\label{fig:S-prime} The spectrum of $H_3$ in slab geometry inside the gapless phase S$'$, $m=1$, $t_1=1.9$, $L=50$. The continuum bands form a single surface/membrane. Two edge modes transverse the hole of the membrane, crossing $E=0$ at $k_y=0$. Both of them are localized on the left edge. }
\end{figure}

A large area of the phase diagram Fig. \ref{fig:phase_diagrams}(c) is occupied by the gapless phase S$'$. An example of the slab spectrum within this phase is shown in Fig. \ref{fig:S-prime} for $t_1=1.9$ and $m=1$. One observes that the two bands merge into a single membrane in the space of $(k_y, \mathrm{Re}E, \mathrm{Im}E)$. While the spectrum on the complex energy plane appears gapless, the membrane possesses a hole around $E=0$, which becomes apparent when projected on the $(k_y,\mathrm{Im}E)$ plane. In other words, within a certain interval of $k_y$ values around $k_y=0$, the spectrum is gapped. This situation is very different from Dirac semimetals, where the gap closes at isolated point degeneracies.  A pair of edge modes $E^{\pm}_{\text{edge}}=\pm \sin k_y$ transverse the hole, and both of them reside on the left edge as proved in Sec. V.B. 
Note the edge states here differ from those in phase S of model $H_2$ which cross $E=0$ at $k_y=\pi$ instead, see Fig. \ref{fig:s_phase}.
We have checked that the edge states are robust against on-site disorder, e.g., in the value of $m$.

In short, this gapless phase is a rather unique feature of non-Hermitian Chern insulators. The edge states are separated from
the continuum bands in the imaginary part of the energy, and therefore in principle can be probed by dynamics \cite{PhysRevB.106.094305}. In some sense, the existence of phase S$'$ attests to the resilience of 
non-Hermitian Chern insulators. When the gap is forced to close, for example, by increasing the non-reciprocal hopping parameter $t_1$ at fixed $m$, the edge modes may survive. The persistence is most apparent along the line of $m=1$ in the phase diagram Fig. \ref{fig:phase_diagrams}(c).

\section{\label{sec:conc}Comparison to earlier work and outlook}

A large body of work has been devoted to study non-Hermitian tight-binding models. For a more comprehensive review of recent progress in this field, see, for example, Refs. \cite{doi:10.1080/00018732.2021.1876991,RevModPhys.93.015005,https://doi.org/10.48550/arxiv.2205.10379,Ghatak_2019}. 
Here we only mention a few works that provided crucial techniques used in our paper or set the stage for our work.
Non-Hermitian Hamiltonians describing particles hopping in 1D, such as the Hatano-Nelson model \cite{PhysRevLett.77.570,PhysRevX.8.031079}, the generalized Su-Schrieffer-Heeger \cite{PhysRevLett.42.1698,PhysRevB.97.045106,PhysRevLett.121.086803,PhysRevLett.121.026808,PhysRevLett.116.133903}, and Rice-Mele model \cite{PhysRevLett.49.1455,PhysRevA.98.042120,Zhang_2019,PhysRevLett.125.186802} are well understood. These models showcase a number of phenomena including the non-Hermitian skin effect and exceptional points in the energy spectrum that are unique to non-Hermitian systems. To characterize the topology of these non-Hermitian systems, unique concepts and techniques were developed, beyond the established framework for Hermitian Bloch Hamiltonians. The initial theoretical efforts focused on the classification of the topological phases based on the dichotomy between point gaps and line gaps \cite{PhysRevX.8.031079,PhysRevX.9.041015,PhysRevLett.123.066405,PhysRevB.99.235112,PhysRevB.100.144106,PhysRevB.99.125103}. Later works gave a more general classification using braid groups \cite{PhysRevB.101.205417,PhysRevB.103.155129} and knots \cite{PhysRevLett.126.010401} for non-Hermitian models with separable bands \cite{PhysRevLett.120.146402}. For multiband systems in 1D with an odd number of bands, invariants can be constructed through the Majorana stellar representation \cite{PhysRevB.104.195131,PhysRevB.101.205309,Xu_2020}.
To restore the bulk boundary correspondence, 
the notion of GBZ was introduced in \cite{PhysRevLett.121.086803} and \cite{PhysRevLett.123.066404} for 1D non-Hermitian Hamiltonians. And the topological origin of the non-Hermitian skin effect in 1D was clarified in Refs. \cite{PhysRevLett.125.126402,PhysRevLett.124.086801,PhysRevX.8.031079,PhysRevLett.124.056802,PhysRevLett.125.126402,PhysRevLett.124.086801,PhysRevLett.122.237601,PhysRevB.100.054301} and attributed to the existence of point gap, which allows the winding number to be defined as the topological invariant. 
Compared to the thorough understanding achieved in 1D, non-Hermitian topological phases in 2D and 3D
are much less understood with many questions remaining open. 

Now we compare our approach and results to a few existing works on non-Hermitian Chern insulators in 2D.
In Ref. \cite{PhysRevLett.121.136802}, Yao \textit{et al.} considered a generalized Qi-Wu-Zhang model with imaginary magnetic fields. They compared the bulk phase diagram (Fig. 1 in Ref. \cite{PhysRevLett.121.136802}) with that of the slab phase diagram (Fig. 3 in Ref. \cite{PhysRevLett.121.136802}) obtained by defining a non-Bloch Chern number. 
These authors treated the non-Hermitian term as a small perturbation, and computed the Chern number using a continuum approximation. In our work, no approximation or extrapolation of the Hamiltonian was made, and the procedure used to compute the GBZ$_s$ and continuum bands are general.  We stress that our strategy of computing the GBZ$_s$ of 2D models builds on the original algorithm outlined in \cite{PhysRevLett.123.066404}, the notion of auxiliary GBZ curves \cite{PhysRevLett.125.226402}, and the self-intersection method \cite{PhysRevB.105.045422}. 

Model $H_2$ in our work was introduced by Kawabata \textit{et al.} \cite{PhysRevB.98.165148}. These authors obtained the slab phase diagram (Fig. 7 in Ref. \cite{PhysRevB.98.165148})
numerically and compared to the bulk phase diagram (Fig. 1 in Ref. \cite{PhysRevB.98.165148}). They also analytically derived the dispersion of the edge modes, and found their localization in the slab geometry (roughly speaking the content of Theorems 3 and 4 here). Here, we take several steps further to obtain the GBZ, the continuum bands, the Chern numbers, and the analytical forms of all the phase boundaries. We also give a precise identification of the gapless phase S and phase C$_2$ in terms of their continuum band structure and Chern numbers. Our phase diagram Fig. \ref{fig:phase_diagrams}(b) labels the phases
differently from \cite{PhysRevB.98.165148}.
These new results, summarized in Theorems 1 and 2 and 5-8, give a thorough understanding of this non-Hermitian Chern insulator.

Other theoretical approaches have been proposed to describe non-Hermitian topological phases in 2D. 
References \cite{PhysRevB.99.245116} introduced a framework based on the transfer matrix in real space to analyze the Qi-Wu-Zhang model with imaginary fields, and Ref. \cite{PhysRevLett.124.056802} employed single and doubled Green's functions to describe the Qi-Wu-Zhang model in an imaginary magnetic field, including the boundary modes and the phase diagram. Reference \cite{PhysRevResearch.2.033069}
employed the entanglement spectrum
to determine topological properties in the gapped phases of the Qi-Wu-Zhang model in an imaginary magnetic field along the $y$ direction. Reference \cite{PhysRevLett.123.246801} constructed real-space topological invariants
to characterize the topological phases for the Qi-Wu-Zhang model in an imaginary magnetic field. Reference \cite{doi:10.1142/S0217979220501465} characterized a non-Hermitian Qi-Wu-Zhang model obtained by a similarity transformation, and proposed a topological invariant for classification.
In passing, we also mention Ref. \cite{PhysRevLett.118.040401} which focused on non-Hermitian Dirac Hamiltonians with gapless spectrum and exceptional points. Reference \cite{PhysRevLett.129.086601} 
proposed an alternative avenue toward realizing non-Hermitian 2D models
using waves backscattered from the boundaries of insulators.
A geometric visualization of the topology of non-Hermitian 2D modes based on the $\mathbf{d}$ vector was advocated in Ref. \cite{PhysRevB.100.075403}. We borrow this perspective in our treatment of model $H_1$. Note, however, the 2D model studied in Eqs. (31) and (32) of Ref. \cite{PhysRevB.100.075403} was more complicated than $H_{1,2,3}$ here.
 A more general version of $H_3$ was mentioned in Ref. \cite{Zhang2022} in discussing the non-Hermitian skin effect.

The main objective of this paper is to outline an algebraic procedure to reliably predict the fascinating slab phase diagrams, including the behaviors of edge modes, for non-Hermitian Chern insulators. The algebraic  procedure does not rely on numerical diagonalization of finite size systems, and therefore is free from the numerical errors that plague the diagonalization of large non-Hermitian matrices. This is not a trivial task, for we have seen GBZ$_s$ with cusps and singularities, topological gapless phases such as S and S$'$ or the higher Chern number phase C$_2$ that are unexpected from bulk analysis, and edge states switching sides while the Chern numbers remain the same.  The breakdown and resurrection of the bulk-edge correspondence is illustrated by two examples, $H_2$ and $H_3$. Such refinement in the understanding of generalized Qi-Wu-Zhang model is achieved by combining various bits of technology available in the literature: analytical continuation, calculation of GBZ curves, analytical solution of the edge spectrum, visualization of the Chern number etc. We hope these examples are helpful to readers who are interested in analyzing other non-Hermitian topological phases of matter in 2D and 3D. 

We have focused exclusively on the slab geometry to limit the paper to a reasonable length. An open question is to analyze the edge and corner modes in finite systems with open boundaries in both the $x$ and $y$ directions, e.g., a rectangle of size $L_x\times L_y$. As pointed out in Ref. \cite{PhysRevB.106.094305}, the edge states of a non-Hermitian Chern insulator may gravitate to corners due to the skin effect, forming the so-called boundary-skin mode. 
The 1D theory established in Refs. \cite{PhysRevLett.125.126402,PhysRevLett.124.086801} can be applied to the effective Hamiltonian that describes the edge degrees of freedom in the slab geometry to understand their corner localization in rectangle geometry. Our preliminary analysis indicates that this scenario is possible for both model $H_2$ and model $H_3$. A comprehensive analysis of the non-Hermitian skin effect in 2D Chern insulators is beyond the scope of this paper and left for future work. 

Non-Hermitian lattice models have been realized in experiments using topological electric circuits \cite{Lee2018,PhysRevLett.126.215302,PhysRevApplied.13.014047}, coupled optical ring resonators \cite{PhysRevB.92.094204,doi:10.1126/science.abf6568,Wang2021}, nitrogen-vacancy centers \cite{doi:10.1126/science.aaw8205,PhysRevLett.126.170506,PhysRevLett.127.090501,Yu2022}, cavity opto-mechanical systems \cite{Patil2022}, phononic crystals with active acoustic components \cite{PhysRevLett.129.084301}, and mechanic metamaterials \cite{doi:10.1073/pnas.2010580117} to name just a few. 
These experimental techniques can potentially be applied to realize the models described here.  Once their topological properties are characterized and understood, non-Hermitian systems may offer exciting opportunities for applications such as topological lasing \cite{doi:10.1126/science.1258004,Brandstetter2014,doi:10.1126/science.aar4003,doi:10.1126/science.aar4005,PhysRevLett.120.113901,Zhao2018,OtaTakataOzawaAmoJiaKanteNotomiArakawaIwamoto+2020+547+567,PhysRevResearch.2.022051,Sone2020,Zapletal:20,Kim2020,PhysRevResearch.3.033042,PhysRevB.104.L140306,PhysRevResearch.4.013195}, enhanced quantum sensing \cite{PhysRevLett.125.180403,PhysRevA.103.042418,PhysRevResearch.4.013131}, and quantum batteries \cite{https://doi.org/10.48550/arxiv.2203.09497}.

\acknowledgments
This work is sponsored by AFOSR Grant No. FA9550- 16-1-0006 and NSF Grant No. PHY- 2011386.
EZ is grateful to Haiping Hu and Bo Liu for helpful discussions. 

\bibliography{chernbib}

\end{document}